\documentclass[10pt, twocolumn]{IEEEtran}
\ifCLASSINFOpdf
\else
\fi
\usepackage{amsfonts,amsmath,amssymb,graphicx,epstopdf,cite,subfigure}

\newtheorem{theorem}{Theorem}[section]
\newtheorem{lemma}[theorem]{Lemma}

\newtheorem{remark}[theorem]{Remark}

\begin{document}
%
\title{Multipath Matching Pursuit}




\author{
Suhyuk (Seokbeop) Kwon,~\IEEEmembership{Student Member,~IEEE,}
Jian Wang,~\IEEEmembership{Student Member,~IEEE,}
        and Byonghyo~Shim,~\IEEEmembership{Senior Member,~IEEE}
\thanks{Copyright (c) 2013 IEEE. Personal use of this material is permitted. However, permission to use this material for any other purposes must be obtained from the IEEE by sending a request to pubs-permissions@ieee.org.}
\thanks{S. Kwon and B. Shim are with School of Information and Communication, Korea University, Seoul, Korea (email: \{shkwon,bshim\}@isl.korea.ac.kr). J. Wang was with School of Information and Communication, Korea University. He is now with Dept. of Statistics, Rutgers University, NJ, USA.}
\thanks{This research was funded by the MSIP (Ministry of Science, ICT \& Future Planning), Korea in the ICT R\&D Program 2013 (No. 1291101110-130010100) and the National Research Foundation of Korea (NRF) grant funded by the Korea government (MEST) (No. 2013056520).}
\thanks{This paper was presented in part at International Symposium on Information Theory (ISIT), Istanbul, July 2013.}
}

\markboth{To appear in IEEE Transactions on Information Theory}%
{Shell \MakeLowercase{\textit{et al.}}: Bare Demo of IEEEtran.cls for Journals}
%



\IEEEtitleabstractindextext{%
\begin{abstract}
In this paper, we propose an algorithm referred to as multipath matching pursuit (MMP) that investigates multiple promising candidates to recover sparse signals from compressed measurements.
Our method is inspired by the fact that the problem to find the candidate that minimizes the residual is readily modeled as a combinatoric tree search problem and the greedy search strategy is a good fit for solving this problem.
In the empirical results as well as the restricted isometry property (RIP) based performance guarantee, we show that the proposed MMP algorithm is effective in reconstructing original sparse signals for both noiseless and noisy scenarios.
\end{abstract}

}

\maketitle

\IEEEdisplaynontitleabstractindextext

%
\IEEEpeerreviewmaketitle

\section{Introduction}
In recent years, compressed sensing (CS) has received much attention as a means to reconstruct sparse signals from compressed measurements \cite{candes2005decoding, candes2006robust, liu2010orthogonal, tropp2007signal, davenport2010analysis, cai2011omp, tong2011sparse, baraniuk2008simple}.
Basic premise of CS is that the sparse signals $\mathbf{x} \in \mathbb{R}^n$ can be reconstructed from the compressed measurements $\mathbf{y} = \mathbf{\Phi} \mathbf{x} \in \mathbb{R}^m$ even when the system representation is underdetermined ($m < n$), as long as the signal to be recovered is sparse (i.e., number of nonzero elements in the vector is small).
%
The problem to reconstruct an original sparse signal is well formulated as an $\ell_0$-minimization problem and $K$-sparse signal $\mathbf{x}$ can be accurately reconstructed using $m = 2K$ measurements in a noiseless scenario \cite{candes2006robust}. 
Since the $\ell_0$-minimization problem is NP-hard and hence not practical, early works focused on the reconstruction of sparse signals using the $\ell_1$-norm minimization technique 
(e.g., basis pursuit \cite{candes2006robust}).

Another line of research, designed to further reduce the computational complexity of the basis pursuit (BP), is the greedy search approach.
In a nutshell, greedy algorithms identify the support (index set of nonzero elements) of the sparse vector $\mathbf{x}$ in an iterative fashion, generating a series of locally optimal updates.
In the well-known orthogonal matching pursuit (OMP) algorithm, the index of column that is best correlated with the modified measurements (often called residual) is chosen as a new element of the support in each iteration \cite{tropp2007signal}.
Therefore, it is not hard to observe that if at least one incorrect index is chosen in the middle of the search, the output of OMP will be simply incorrect.
In order to mitigate the algorithmic weakness of OMP, modifications of OMP, such as inclusion of thresholding (e.g., StOMP \cite{donoho2006sparse}), selection of indices exceeding the sparsity level followed by a pruning (e.g., CoSaMP \cite{needell2009cosamp} and SP \cite{dai2009subspace}), and multiple indices selection (e.g., gOMP \cite{wang2012gomp}), have been proposed.
These approaches are better than OMP in empirical performance as well as theoretical performance guarantee, but their performance in the noisy scenario is far from being satisfactory, especially when compared to the best achievable bound obtained from Oracle estimator.\footnote{The estimator that has prior knowledge on the support is called Oracle estimator.}

\begin{figure*}
	\begin{center}
		\includegraphics[width=170mm]{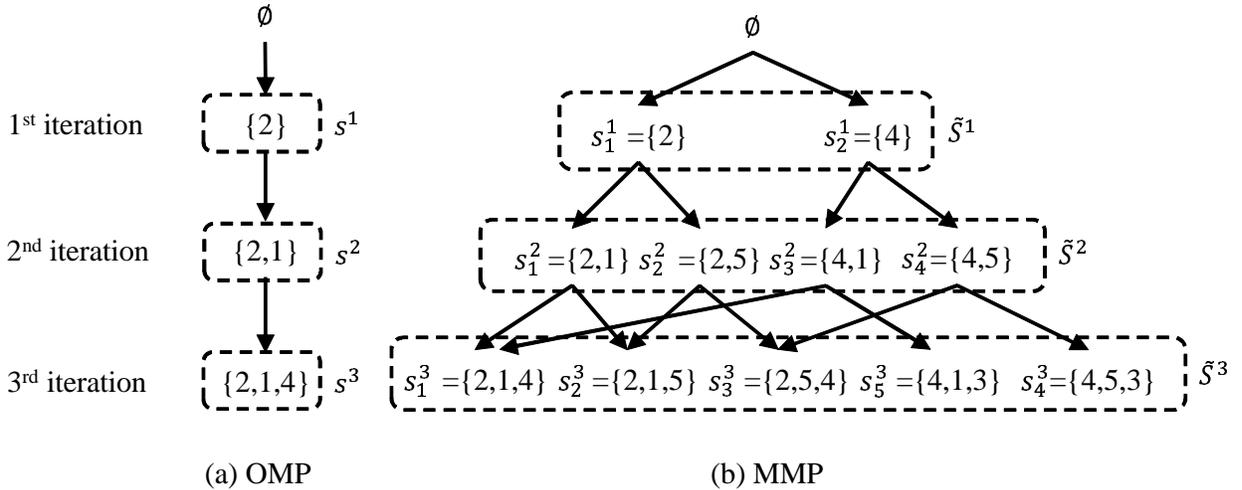}
		\caption{Comparison between the OMP and the MMP algorithm ($L=2$ and $K=3$).}
		\label{fig:concept_tree}
	\end{center}
\end{figure*}

The main goal of this paper is to go further and pursue a smart grafting of two seemingly distinct principles: {\it combinatoric approach} and {\it greedy algorithm}.
Since all combinations of $K$-sparse indices can be interpreted as candidates in a tree (see Fig. \ref{fig:concept_tree}) and each layer of the tree can be sorted by the magnitude of the correlation between the column of sensing matrix and residual, the problem to find the candidate that minimizes the residual is readily modeled as a combinatoric tree search problem.
Note that in many disciplines, the tree search problem is solved using an efficient search algorithm, not by the brute-force enumeration.
Well-known examples include Viterbi decoding for maximum likelihood (ML) sequence detection \cite{viterbi1967error}, sphere decoding for ML detection \cite{viterbo1999universal, bshim2008sphere},
and list sphere decoding for maximum a posteriori (MAP) detection \cite{hochwald2003achieving}.
Some of these return the optimal solution while others return an approximate solution, but the common wisdom behind these algorithms is that they exploit the structure of tree to improve the search efficiency.
%
%

In fact, the proposed algorithm, henceforth referred to as {\it multipath matching pursuit} (MMP), performs the tree search with the help of the greedy strategy.
%
Although each candidate brings forth multiple children and hence the number of candidates increases as an iteration goes on, the increase is actually moderate since many candidates are overlapping in the middle of search (see Fig. \ref{fig:concept_tree}).
Therefore, while imposing reasonable computational overhead, the proposed method achieves considerable performance gain over existing greedy algorithms.
In particular, when compared to the variations of OMP which in essence trace and output the single candidate, MMP examines multiple full-blown candidates and then selects the final output in the last minutes so that it improves the chance of selecting the true support substantially.

The main contributions of this paper are summarized as follows:
\begin{itemize}
	\item We present a new sparse signal recovery algorithm, termed MMP, for pursuing efficiency in the reconstruction of sparse signals.
		Our empirical simulations show that the recovery performance of MMP is better than the existing sparse recovery algorithms, in both noiseless and noisy scenarios.
	\item We show that the perfect recovery of any K-sparse signal can be ensured by the MMP algorithm in the noiseless scenario if the sensing matrix satisfies the restricted isometry property (RIP) condition $\delta_{L+K} < \frac{\sqrt{L}}{\sqrt{K}+2\sqrt{L}}$ where $L$ is the number of child paths for each candidate (Theorem \ref{thm:sc_final}).
		In particular, if $L = K$, the recovery condition is simplified to $\delta_{2K} < 0.33$.
		This result, although slightly worse than the condition of BP ($\delta_{2K} < \sqrt{2}-1$), is fairly competitive among conditions of state of the art greedy recovery algorithms (Remark \ref{cor:cond_LK}).
	\item We show that the true support is identified by the MMP algorithm if $\min_{x \neq 0} |x| \geq c \|\mathbf{v}\|_2$ where $\mathbf{v}$ is the noise vector and $c$ is a function of $\delta_{2K}$ (Theorem \ref{thm:sc_gen_n}).
		Under this condition, which in essence corresponds to the high signal-to-noise ratio (SNR) scenario, we can remove all non-support elements and columns associated with these so that we can obtain the best achievable system model $\mathbf{y} = \mathbf{\Phi}_T \mathbf{x}_T + \mathbf{v}$ where $\mathbf{\Phi}_T$ and $\mathbf{x}_T$ are the sensing matrix and signal correspond to the true support $T$, respectively (refer to the notations in the next paragraph).
		Remarkably, in this case, the performance of MMP becomes equivalent to that of the least square (LS) method of the overdetermined system (often referred to as Oracle LS estimator \cite{chan2013projection}).
Indeed, we observe from empirical simulations that MMP performs close to the Oracle LS estimator in the high SNR regime.
	\item We propose a modification of MMP, referred to as depth-first MMP (MMP-DF), for strictly controlling the computational complexity of MMP.
		When combined with a properly designed search strategy termed {\it modulo strategy}, MMP-DF performs comparable to the original (breadth-first) MMP while achieving substantial savings in complexity.
\end{itemize}

The rest of this paper is organized as follows.
In Section~\ref{sec:algorithm}, we introduce the proposed MMP algorithm.
In Section~\ref{sec:mainresults}, we analyze the RIP based condition of MMP that ensures the perfect recovery of sparse signals in noiseless scenario.
In Section~\ref{sec:noisy_anal}, we analyze the RIP based condition of MMP to identify the true support from noisy measurements.
In Section~\ref{sec:complxt}, we discuss a low-complexity implementation of the MMP algorithm (MMP-DF).
In Section~\ref{sec:sim}, we provide numerical results and conclude the paper in Section~\ref{sec:con}.

We briefly summarize notations used in this paper.
$x_i$ is the $i$-th element of vector $\mathbf{x}$.
$I_1-I_2 = I_1 \setminus (I_1 \cap I_2 )$ is the set of all elements contained in $I_1$ but not in $I_2$.
$|\Lambda|$ is the cardinality of $\Lambda$.
$\mathbf{\Phi}_{\Lambda} \in \mathbb{R}^{m \times |\Lambda|}$ is a submatrix of $\mathbf{\Phi}$ that contains columns indexed by $\Lambda$. For example, if $\mathbf{\Phi} = [\phi_1 ~\phi_2 ~\phi_3 ~\phi_4]$ and $\Lambda = \{ 1, 3\}$, then $\mathbf{\Phi}_{\Lambda} = [\phi_1 ~\phi_3]$.
Let $\Omega=\{1, 2, \ldots, n \}$ be the column indices of matrix $\mathbf{\Phi}$, then $T= \{ i \mid i \in \Omega, \text{ } x_i \neq 0 \}$ and $T^C= \{ j \mid j \in \Omega, \text{ } x_j = 0 \}$ denote the support of vector $\mathbf{x}$ and its complement, respectively.
$s^k_i$ is the $i$-th candidate in the $k$-th iteration and
%
$S^k = \{ s^k_1, s^k_2, \ldots, s^k_u \}$ is the set of candidates in the $k$-th iteration.
%
%
$\Omega^k$ is a set of all possible combinations of $k$ columns in $\mathbf{\Phi}$.
For example, if $\Omega = \left \{ 1, 2, 3\right \}$ and $k=2$, then $\Omega^k=\left \{ \left \{ 1,2\right \}, \left \{ 1,3\right \}, \left \{ 2,3\right \}\right \}$.
$\mathbf{\Phi}'$ is a transpose matrix of $\mathbf{\Phi}$.
If $\mathbf{\Phi}$ is full column rank, then $\mathbf{\Phi}^{\dagger} = \left ( \mathbf{\Phi}'\mathbf{\Phi} \right )^{-1} \mathbf{\Phi}'$ is the Moore-Penrose pseudoinverse of $\mathbf{\Phi}$.
$\mathbf{P}_{\Lambda} = \mathbf{\Phi}_{\Lambda}\mathbf{\Phi}_{\Lambda}^{\dagger}$ and $\mathbf{P}_{\Lambda}^\bot  = \mathbf{I} - \mathbf{P}_{\Lambda}$ are the projections onto $span(\mathbf{\Phi}_{\Lambda})$ and the orthogonal complement of $span(\mathbf{\Phi}_{\Lambda})$, respectively.
%

\section{MMP Algorithm}\label{sec:algorithm}

Recall that the $\ell_0$-norm minimization problem to find out the sparsest solution of an underdetermined system is given by
\begin{IEEEeqnarray}{rCl}
\label{eq:basic2} \min_{\mathbf{x} } \|\mathbf{x}\|_{{0}}
\hspace{0.7cm} \mbox{subject to} \hspace{0.3cm}\mathbf{\Phi x = y}.
\end{IEEEeqnarray}
%
In finding the solution of this problem, all candidates (combination of columns) satisfying the equality constraint should be tested.
In particular, if the sparsity level and the signal length are set to $K$ and $n$, respectively, then $n \choose K$ candidates should be investigated, which is obviously prohibitive for large $n$ and nontrivial $K$ \cite{candes2005decoding}.
In contrast, only one candidate is searched heuristically in the OMP algorithm \cite{tropp2007signal}.
Although OMP is simple to implement and also computationally efficient, due to the selection of the single candidate in each iteration, it is very sensitive to the selection of index.
As mentioned, the output of OMP will be simply wrong if an incorrect index is chosen in the middle of the search.
In order to reduce the chance of missing the true index and choosing incorrect one,
various approaches investigating {\it multiple indices} have been proposed.
In \cite{donoho2006sparse}, StOMP algorithm identifying more than one indices in each iteration was proposed.
In this approach, indices whose magnitude of correlation exceeds a deliberately designed threshold are chosen \cite{donoho2006sparse}.
In \cite{needell2009cosamp} and  \cite{dai2009subspace}, CoSaMP and SP algorithms maintaining $K$ support elements in each iteration were introduced.
In \cite{wang2012gomp}, another variation of OMP, referred to as generalized OMP (gOMP), was proposed.
By choosing multiple indices corresponding to $N (> 1)$ largest correlation in magnitude in each iteration, gOMP reduces the misdetection probability at the expense of increase in the false alarm probability.
%

\begin{table*}[!t]
\centering
\caption{The MMP algorithm}
\label{tab:alg1}
\begin{tabular}{l r}
\hline
\hline
\multicolumn{2}{l}{\textbf{Input:} measurement $\mathbf{y}$, sensing matrix $\mathbf{\Phi}$, sparsity $K$, number of path $L$ }\\
\multicolumn{2}{l}{\textbf{Output:} estimated signal $\hat{\mathbf{x}}$	}\\
\multicolumn{2}{l}{\textbf{Initialization:} $k:=0$ (iteration index), $\mathbf{r}^0 := \mathbf{y}$ (initial residual), $S^0 := \left \{ \emptyset \right \}$		 }\\
\hline
\multicolumn{2}{l}{{\bf while} {$k < K$} {\bf do}										}\\
\multicolumn{2}{l}{	\hspace{4mm}$k :=k+1$, $u :=0$, $S^{k} :=\emptyset$				 }\\
\multicolumn{2}{l}{	\hspace{4mm}{\bf for} {$ i = 1$ {\bf to} $| S^{k-1} | $} {\bf do}			}\\
					\hspace{8mm}	$ \tilde{\pi} := \arg \mathop {\max} \limits_{ \left | \pi \right | = L} { \| ( \mathbf{\Phi}' \mathbf{r}^{k-1}_i )_{\pi} \|_2^2 } $ & ({\it choose $L$ best indices})	\\
\multicolumn{2}{l}{	\hspace{8mm}	{\bf for} {$ j=1 $ {\bf to} $ L $} {\bf do}													 }\\
					\hspace{12mm}		$s_{temp} := s^{k-1}_i \cup \left \{ \tilde{\pi}_j \right \}$	& ({\it construct a temporary path})\\
					\hspace{12mm}		{\bf if} {$s_{temp} \not \in S^{k}$}	{\bf then} & ({\it check if the path already exists})		\\
					\hspace{16mm}			$u :=u+1$							& ({\it candidate index update})		 \\
                    \hspace{16mm}			$s^k_u :=s_{temp}$		& ({\it path update})
\\
					\hspace{16mm}			$S^{k} := S^{k} \cup \{ s^{k}_u \}$	& ({\it update the set of path})\\
					\hspace{16mm}			$\hat{\mathbf{x}}^k_u := \mathbf{\Phi}^{\dagger}_{s^{k}_u} \mathbf{y}$	& ({\it perform estimation})	 \\
					\hspace{16mm}			$\mathbf{r}^k_u := \mathbf{y}-\mathbf{\Phi}_{s^{k}_u} \hat{\mathbf{x}}^k_u$	 & ({\it residual update})\\
\multicolumn{2}{l}{	\hspace{12mm}		{\bf end if}																	 }\\
\multicolumn{2}{l}{	\hspace{8mm}	{\bf end for}																		 }\\
\multicolumn{2}{l}{	\hspace{4mm}{\bf end for}																			 }\\
\multicolumn{2}{l}{{\bf end while}																							 }\\
					{$ u^* := \arg \mathop {\min} { \| \mathbf{r}_{u}^K \|_2^2 } $}		& ({\it find index of the best candidate})		\\
\multicolumn{2}{l}{{$ s^* := s^K_{u^*}$}					 }\\
\multicolumn{2}{l}{{\bf return} $\hat{\mathbf{x}} = \mathbf{\Phi}^{\dagger}_{s^*} \mathbf{y}$															 }\\
\hline
\end{tabular}
\end{table*}

\begin{figure}
\begin{center}
\ifCLASSOPTIONonecolumn
	\includegraphics[width=140mm]{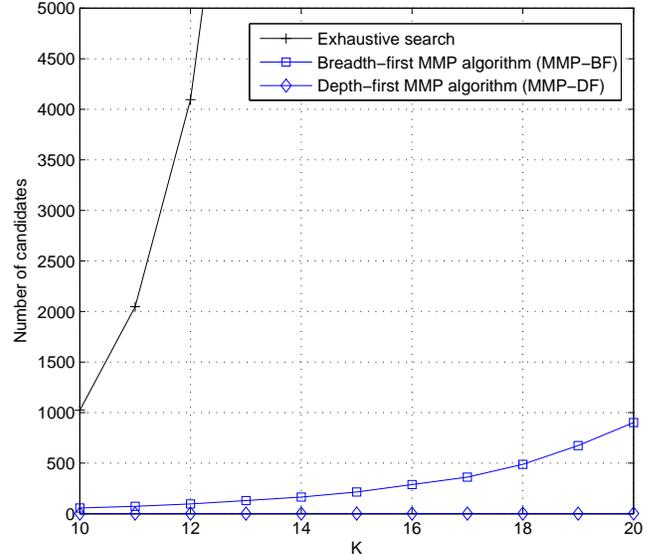}
\else
	\includegraphics[width=95mm]{2K_comp}
\fi
	\caption{The number of candidates for $K$-sparse signal recovery (averages of 1000 trials with $L=2$). The MMP algorithm in this section is considered as the breadth-first MMP. In Section \ref{sec:complxt}, we present a low-complexity MMP algorithm based on the depth-first search (MMP-DF).}
	\label{fig:2k_comp}
\end{center}
\end{figure}

While these approaches exploit {\it multiple indices} to improve the reconstruction performance of sparse signals, they maintain a single candidate in the search process. Whereas, the proposed MMP algorithm searches {\it multiple promising candidates} and then chooses one minimizing the residual in the final moment.
In this sense, one can think of MMP as an approximate algorithm to find the candidate minimizing the cost function $J(\mathbf{x})=\| \mathbf{y} - \mathbf{\Phi} \mathbf{x} \|_2$ subject to the sparsity constraint $\left \| \mathbf{x} \right \|_0 = K$.
Due to the investigation of multiple promising full-blown candidates, MMP improves the chance of selecting the true support.
In fact, noting that the effect of the random noise vector cannot be accurately judged by just looking at the partial candidate, and more importantly, incorrect decision affects subsequent decision in many greedy algorithms,\footnote{This phenomenon is similar to the error propagation problem in the successive interference cancellation \cite{verdu1998mimo}} it is not hard to convince oneself that the strategy to investigate multiple candidates is effective in noisy scenario.
We compare operations of the OMP and MMP algorithm in Fig. \ref{fig:concept_tree}.
While a single path (candidate) is maintained for all iterations of the OMP algorithm, each path generates $L$ child paths in the MMP algorithm.\footnote{In this paper, we use path and candidate interchangeably.}
In the $k$-th iteration, $L$ indices of columns $\tilde{\pi}_1, \cdots \tilde{\pi}_L$ that are maximally correlated with the residual\footnote{The residual is expressed as $\mathbf{r}^{k-1}_i = \mathbf{y} - \mathbf{\Phi}_{s^{k-1}_i} \hat{\mathbf{x}}_{s^{k-1}_i}$ where $s^{k-1}_i$ is the $i$-th candidate in the $(k-1)$-th iteration and $\hat{\mathbf{x}}_{s^{k-1}_i}$ is the estimate of $\hat{\mathbf{x}}$ using columns indexed by $s^{k-1}_i$.} become new elements of the child candidates (i.e., $ \{\tilde{\pi}_1, \cdots \tilde{\pi}_L \} = \arg \mathop {\max} \limits_{ \left | \pi \right | = L} { \| ( \mathbf{\Phi}' \mathbf{r}^{k-1}_i )_{\pi} \|_2^2 } $).
At a glance, it seems that the number of candidates increases by the factor of $L$ in each iteration, resulting in $L^K$ candidates after $K$ iterations.
In practice, however, the number of candidates increases moderately since large number of paths is overlapping during the search.
As illustrated in the Fig. \ref{fig:concept_tree}(b), $s^3_3=\{ 2, 5, 4\}$ is the child of $s^2_2=\{ 2, 5\}$ and $s^2_4=\{ 4, 5\}$ and $s^3_1=\{2, 1, 4\}$ is the child of $s^2_1=\{2, 1\}$ and $s^2_3=\{4, 1\}$ so that the number of candidates in the second iteration is $4$ but that in the third iteration is just $5$.
Indeed, as shown in Fig. \ref{fig:2k_comp}, the number of candidates of MMP is much smaller than that generated by the exhaustive search.
In Table \ref{tab:alg1}, we summarize the proposed MMP algorithm.
%

\section{Perfect Recovery Condition for MMP} \label{sec:mainresults}

In this section, we analyze a recovery condition under which MMP can accurately recover $K$-sparse signals in the noiseless scenario.
Overall, our analysis is divided into two parts.
In the first part, we consider a condition ensuring the successful recovery in the initial iteration ($k = 1$).
In the second part, we investigate a condition guaranteeing the success in the non-initial iteration ($k > 1$).
By success we mean that an index of the true support $T$ is chosen in the iteration.
By choosing the stricter condition between two as our final recovery condition, the perfect recovery condition of MMP can be identified.	\label{rev_page:7}

In our analysis, we use the RIP of the sensing matrix $\mathbf{\Phi}$.
A sensing matrix $\mathbf{\Phi}$ is said to satisfy the RIP of order $K$ if there exists a constant $\delta(\mathbf{\Phi}) \in (0,1)$ such that
\begin{equation}
	(1-\delta(\mathbf{\Phi})) \| \mathbf{x} \|_2^2 \leq \| \mathbf{\Phi}\mathbf{x} \|_2^2 \leq (1+\delta(\mathbf{\Phi})) \| \mathbf{x} \|_2^2		 \label{eq:rip}
\end{equation}
for any $K$-sparse vector $\mathbf{x}$.
In particular, the minimum of all constants $\delta (\mathbf{\Phi})$ satisfying \eqref{eq:rip} is called the restricted isometry constant $\delta_K (\mathbf{\Phi})$.
In the sequel, we use $\delta_K$ instead of $\delta_K (\mathbf{\Phi})$ for notational simplicity.
Roughly speaking, we say a matrix satisfies the RIP if $\delta_K$ is not too close to one. Note that if $\delta_K \approx 1$, then it is possible that $\| \mathbf{\Phi}\mathbf{x} \|_2^2 \approx \mathbf{0}$ (i.e., $\mathbf{x}$ is in the nullspace of $\mathbf{\Phi}$) so that the measurements $\mathbf{y} = \mathbf{\Phi}\mathbf{x}$ do not preserve any information on $\mathbf{x}$ and the recovery of $\mathbf{x}$ would be nearly impossible.
On the other hand, if $\delta_K \approx 0$, the sensing matrix is close to orthonormal so that the reconstruction of $\mathbf{x}$ would be guaranteed almost surely.
In many algorithms, therefore, the recovery condition is expressed as an upper bound of the restricted isometry constant (see Remark \ref{cor:cond_LK}).

Following lemmas, which can be easily derived from the definition of RIP, will be useful in our analysis.

\ifCLASSOPTIONonecolumn
\begin{lemma}[Monotonicity of the restricted isometry constant \cite{candes2005decoding}]
\else
\begin{lemma}[Monotonicity of the restricted isometry constant \cite{candes2005decoding}]
\fi
\label{lem:rip_mono}
If the sensing matrix $\mathbf{\Phi}$ satisfies the RIP of both orders $K_1$ and $K_2$, then $\delta _{K_1} \leq \delta _{K_2}$ for any $K_1 \leq K_2$.
\end{lemma}

\begin{lemma} [Consequences of RIP \cite{candes2005decoding}]\label{lem:rip_cons} For $I \subset \Omega$, if $\delta_{\left| I \right|} < 1$ then for any ${\mathbf{x}} \in {\mathbb{R}^{\left| I \right|}}$,
\begin{IEEEeqnarray}{rCl}
\left( 1 - \delta _{\left| I \right|} \right) \left\| \mathbf{x} \right\|_2
\leq \left\| \mathbf{\Phi}_I' \mathbf{\Phi}_I \mathbf{x} \right\|_2
\leq \left( 1 + \delta _{\left| I \right|} \right) \left\| \mathbf{x} \right\|_2,	\\
\frac {1} {1 + \delta _{\left| I \right|}}  \left\| \mathbf{x} \right\|_2
\leq \| \left(\mathbf{\Phi}_I' \mathbf{\Phi}_I \right)^{-1} \mathbf{x}  \|_2
\leq \frac {1} {1 - \delta _{\left| I \right|}} \left\| \mathbf{x} \right\|_2.
\end{IEEEeqnarray}
\end{lemma}

\begin{lemma}[Lemma 2.1 in \cite{candes2008restricted}] \label{lem:rip_add} Let ${I_1}, {I_2}\subset \Omega$ be two disjoint sets ($I_1 \cap I_2 = \emptyset$). If $\delta _{|I_1| +|I_2|} < 1$, then
\begin{equation}
\left\| \mathbf{\Phi}_{I_1}' \mathbf{\Phi}_{I_2} \mathbf{x} \right\|_2 \leq \delta _{|I_1| + |I_2|} \left\| \mathbf{x} \right\|_2
\end{equation}
holds for any $\mathbf{x}$.
\end{lemma}

\begin{lemma} \label{lem:matrix_up}
For $m \times n$ matrix $\mathbf{\Phi}$, $\left\| \mathbf{\Phi} \right\|_2$ satisfies
\begin{equation}
\left\| \mathbf{\Phi} \right\|_2 = \sqrt{\lambda_{\max}(\mathbf{\Phi}'\mathbf{\Phi})} \leq \sqrt{1+\delta_{\min(m,n)}}	 \label{rev_eq:eigen_def}
\end{equation}
where $\lambda_{\max}$ is the maximum eigenvalue.
\end{lemma}

\subsection{Success Condition in the First Iteration}
In the first iteration, MMP computes the correlation between measurements $\mathbf{y}$ and each column $\mathbf{\phi}_i$ of $\mathbf{\Phi}$ and then selects $L$ indices whose column has largest correlation in magnitude.
%
%
Let $\Lambda$ be the set of $L$ indices chosen in the first iteration, then
\begin{equation}
\left\| \mathbf{\Phi}_{\Lambda}' \mathbf{y} \right\|_2
= \mathop {\max} \limits_{\left| I \right| = L } \sqrt{\sum\limits_{i\in I} \left| \left\langle \mathbf{\phi}_{i},\mathbf{y} \right\rangle  \right| ^2}.	\label{eq:sc_init_sel_def}
\end{equation}
Following theorem provides a condition under which at least one correct index belonging to $T$ is chosen in the first iteration.

\begin{theorem}\label{thm:sc_init}
Suppose $\mathbf{x} \in \mathbb{R}^n$ is $K$-sparse signal, then among $L(\leq K)$ candidates at least one contains the correct index in the first iteration of the MMP algorithm if the sensing matrix $\mathbf{\Phi}$ satisfies the RIP with
\begin{equation}
\delta_{K+L} < \frac{\sqrt{L}}{\sqrt{K}+\sqrt{L}}.
\label{eq:sc_init}
\end{equation}
\end{theorem}

\begin{IEEEproof}
From \eqref{eq:sc_init_sel_def}, we have
\begin{IEEEeqnarray}{rCl}
\frac{1}{\sqrt{L}} \left\| \mathbf{\Phi}_{\Lambda}' \mathbf{y} \right\|_{2}
&=& \frac{1}{\sqrt{L}} \mathop {\max} \limits_{\left| I \right| = L } \sqrt{\sum\limits_{i\in I} \left| \left\langle \mathbf{\phi}_i ,\mathbf{y} \right\rangle \right|^2}	\nonumber	\\
& = &  \mathop {\max} \limits_{\left| I \right| = L } \sqrt{\frac{1}{ |I| }\sum\limits_{i\in I} \left| \left\langle \mathbf{\phi}_i ,\mathbf{y} \right\rangle \right|^2}	\nonumber	\\
& \geq &  \sqrt{\frac{1}{|T|} \sum\limits_{i\in T} \left| \left\langle \mathbf{\phi}_i, \mathbf{y} \right\rangle \right|^2}	 \nonumber	\\
& = & \frac{1}{\sqrt{K}} \left\| \mathbf{\Phi}_{T}' \mathbf{y} \right\|_2
\end{IEEEeqnarray}
where $\left| T \right|=K$.
Since $\mathbf{y} = \mathbf{\Phi}_{T} \mathbf{x}_{T}$, we further have
\begin{IEEEeqnarray}{rCl}
 \left\| \mathbf{\Phi}_{\Lambda}' \mathbf{y} \right\|_{2}
&\geq & \sqrt{\frac{ L }{K}} \left\| \mathbf{\Phi}_{T}' \mathbf{\Phi }_{T}\mathbf{x}_{T} \right\|_{2} 	\nonumber	\\
&\geq & \sqrt{\frac{ L }{K}} \left( 1-\delta_{K} \right) \left\| \mathbf{x} \right\|_{2} \label{eq:sc_init_1}
\end{IEEEeqnarray}
where \eqref{eq:sc_init_1} is due to Lemma \ref{lem:rip_cons}.

On the other hand, when an incorrect index is chosen in the first iteration (i.e., $\Lambda \cap T = \emptyset$),
\begin{equation}
\left\| \mathbf{\Phi}_{\Lambda}' \mathbf{y} \right\|_2=\left\| \mathbf{\Phi}_{\Lambda}' \mathbf{\Phi}_{T} \mathbf{x}_{T} \right\|_2
\leq \delta_{K +L} \left\| \mathbf{x} \right\|_{2},
\end{equation}
where the inequality follows from Lemma \ref{lem:rip_add}.
This inequality contradicts \eqref{eq:sc_init_1} if
\begin{equation} \label{eq:sc_init_contra}
\delta_{K +L} \left\| \mathbf{x} \right\|_{2} < \sqrt{\frac{L}{K}}\left( 1-\delta_{K} \right) \left\| \mathbf{x} \right\|_{2}.
\end{equation}
In other words, under \eqref{eq:sc_init_contra} at least one correct index should be chosen in the first iteration ($T^1_i \in \Lambda$).
Further, since $\delta_K  \leq \delta_{K + N}$ by Lemma \ref{lem:rip_mono}, \eqref{eq:sc_init_contra} holds true if
\begin{equation}
\delta_{K + L} \left\| \mathbf{x} \right\| _{2} < \sqrt{\frac{L}{K}}\left( 1-\delta_{K + L} \right)\left\| \mathbf{x} \right\|_{2}.
\end{equation}
Equivalently,
\begin{equation}
\delta_{K + L} < \frac{\sqrt{L}}{\sqrt{K}+\sqrt{L}}.
\end{equation}
In summary, if $\delta_{K + L} <\frac{\sqrt{L}}{\sqrt{K}+\sqrt{L}}$, then among $L$ indices at least one belongs to $T$ in the first iteration of MMP.
\end{IEEEproof}

\subsection{Success Condition in Non-initial Iterations}

Now we turn to the analysis of the success condition for non-initial iterations.
In the $k$-th iteration ($k>1$), we focus on the candidate $s^{k-1}_i$ whose elements are exclusively from the true support $T$ (see Fig. \ref{fig:gen_itr_relation}).
Our key finding is that at least one of $L$ indices generated from $s^{k-1}_i$ is the element of $T$ under $\delta_{K + L} < \frac{\sqrt{L}}{\sqrt{K}+2\sqrt{L}}$.
Formal description of our finding is as follows.

\begin{figure*}
\begin{center}
	\includegraphics[width=150mm]{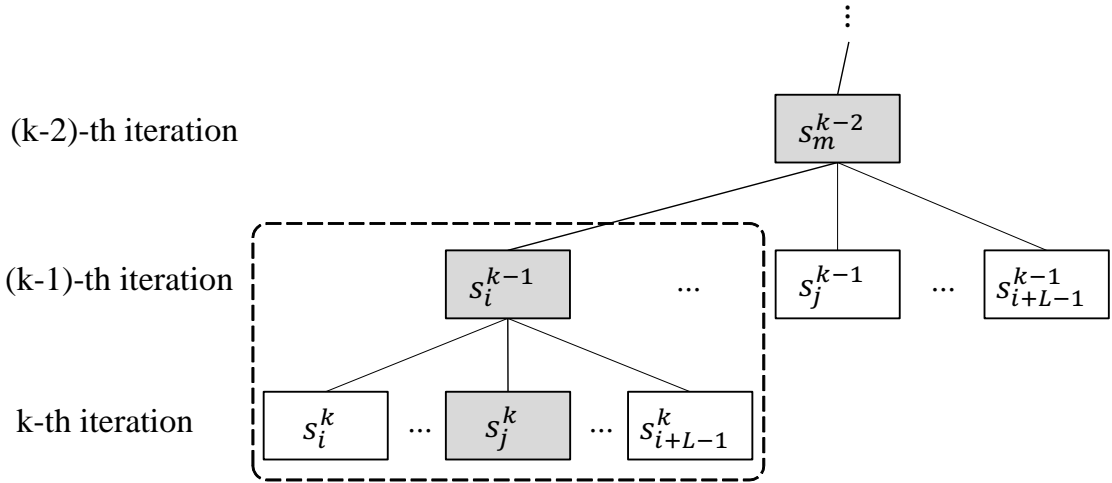}
	\caption{Relationship between the candidates in the $(k-1)$-th iteration and those in the $k$-th iteration. Candidates inside the gray box contain elements of true support $T$ only.}
	\label{fig:gen_itr_relation}
\end{center}
\end{figure*}

%
%
%
%
\begin{theorem}\label{thm:sc_gen}
Suppose a candidate $s^{k-1}_i$ includes indices only in $T$, then among $L$ child candidates at least one chooses an index in $T$ under
\begin{equation}
\delta_{K+L} < \frac{\sqrt{L}}{\sqrt{K}+2\sqrt{L}}.
\label{eq:sc_gen}
\end{equation}
\end{theorem}

\vspace{0.1cm}
Before we proceed, we provide definitions and lemmas useful in our analysis.
Let $f_j$ be the $j$-th largest correlated index in magnitude between $\mathbf{r}^{k-1}_i$ and $\{ \mathbf{\phi}_j \}_{j \in T^C}$ (set of incorrect indices).
That is, $$f_j = \arg \mathop {\max} \limits_{ u \in T^C \setminus \{f_1, \ldots, f_{(j-1)}\}} \left|\left <\mathbf{\phi}_u, \mathbf{r}^{k-1}_i \right> \right|.$$
Let $F_L$ be the set of these indices ($F_L = \left\{ f_1, f_2, \cdots, f_L \right\}$).
Also, let $\alpha_j^k$ be the $j$-th largest correlation in magnitude between $\mathbf{r}^{k-1}_i$ and columns indexed by $f_j$.
That is,
\begin{equation}
\alpha_j^k = \left|\left <\mathbf{\phi}_{f_j}, \mathbf{r}^{k-1}_i \right> \right|.	\label{eq:def_aL}
\end{equation}
Note that $\alpha_j^k$ are ordered in magnitude ($\alpha_1^k$ $\geq$ $\alpha_2^k$ $\geq$ $\cdots$).
Finally, let $\beta_j^k$ be the $j$-th largest correlation in magnitude between $\mathbf{r}^{k-1}_i$ and columns whose indices belong to $T - s^{k-1}_i$ (the set of remaining true indices).
That is,
\begin{equation}
\beta_j^k = \left|\left<\mathbf{\phi}_{\varphi_j}, \mathbf{r}^{k-1}_i\right>\right|	\label{eq:def_b1}
\end{equation}
where $\varphi_j = \arg \mathop {\max} \limits_{ u \in \left( T-s^{k-1} \right)\setminus \{\varphi_1, \ldots, \varphi_{j-1} \}} \left|\left <\mathbf{\phi}_u, \mathbf{r}^{k-1}_i \right> \right|$.
Similar to $\alpha_j^k$, $\beta_j^k$ are ordered in magnitude ($\beta_1^k$ $\geq$ $\beta_2^k$ $\geq$ $\cdots$).
In the following lemmas, we provide an upper bound of $\alpha_L^k$ and a lower bound of $\beta_1^k$, respectively.
\begin{lemma}\label{lem:aL}
Suppose a candidate $s^{k-1}_i$ includes indices only in $T$, then $\alpha_L^k$ satisfies
\begin{equation}
\alpha_L^k \leq \left( \delta _{L+K-k+1} + \frac{\delta _{L+k-1} \delta_{K}} {1 - \delta_{k-1}} \right ) \frac{\left \| \mathbf{x}_{T - s^{k-1}_i} \right \|_2 }{\sqrt{L}}.
\end{equation}
\end{lemma}
\begin{IEEEproof}
See Appendix \ref{app:aL}.
\end{IEEEproof}

\begin{lemma}\label{lem:b1}
Suppose a candidate $s^{k-1}_i$ includes indices only in $T$, then $\beta_1^k$ satisfies
\ifCLASSOPTIONonecolumn
\begin{equation}
\beta_1^k
\geq
\left( 1- \delta_{K-k+1}-\frac{\delta_{K}^2}{1-\delta_{k-1}} \right) \frac{\left\| \mathbf{x}_{T - s^{k-1}_i} \right\|_2}{\sqrt{K-k+1}}.
\end{equation}
\else
\begin{IEEEeqnarray}{rCl}
\beta_1^k
& \geq &
\left( 1- \delta_{K-k+1}-\frac{\sqrt{1+\delta_{K-k+1}} \sqrt{1+\delta_{k-1}} \delta_{K}}{1-\delta_{k-1}} \right) \nonumber	 \\
& & \cdot \frac{\left\| \mathbf{x}_{T - s^{k-1}_i} \right\|_2}{\sqrt{K-k+1}}.
\end{IEEEeqnarray}
\fi

\end{lemma}
\begin{IEEEproof}
See Appendix \ref{app:b1}.
\end{IEEEproof}

\vspace{0.4cm}

\begin{figure*}
\begin{center}
	\includegraphics[width=100mm]{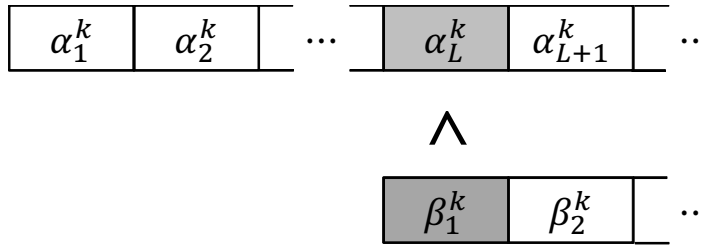}
	\caption{Comparison between $\alpha_N^k$ and $\beta_1^k$.
If $\beta_1^k > \alpha_N^k$, then among $L$ indices chosen in the $K$-th iteration, at least one is from the true support $T$.}
	\label{fig:alpha_beta}
\end{center}
\end{figure*}

\begin{IEEEproof}[\textbf{Proof of Theorem \ref{thm:sc_gen}}]
From the definitions of $\alpha_j^k$ and $\beta_j^k$, it is clear that a sufficient condition under which at least one out of $L$ indices is true in the $k$-th iteration of MMP is (see Fig. \ref{fig:alpha_beta})
\begin{equation}
\beta_1^k > \alpha_L^k
\label{eq:sc_gen_sketch}
\end{equation}
First, from Lemma \ref{lem:rip_mono} and \ref{lem:aL}, we have
\begin{IEEEeqnarray}{rCl}
\alpha_L^k
	&\leq & \left( \delta _{L+K - k+1} + \frac{\delta _{L+k-1} \delta_{K}} {1 - \delta_{k-1}} \right ) \frac{\left \| \mathbf{x}_{T - s^{k-1}_i} \right \|_2 }{\sqrt{L}}	\nonumber	\\
	&\leq & \left( \delta _{L+K} + \frac{\delta _{L+K} \delta_{L+K}} {1 - \delta_{L+K}} \right ) \frac{\left \| \mathbf{x}_{T - s^{k-1}_i} \right \|_2 }{\sqrt{L}}			\nonumber	\\
	&= & \frac{\delta _{L+K}} {1 - \delta_{L+K}} \frac{\left \| \mathbf{x}_{T - s^{k-1}_i} \right \|_2 }{\sqrt{L}}.	 \label{eq:lem_aL_ex}
\end{IEEEeqnarray}
Also, from Lemma \ref{lem:rip_mono} and \ref{lem:b1}, we have
\begin{IEEEeqnarray}{rCl}
	\beta_1^k
	&\geq & \left( 1- \delta_{K-k+1}-\frac{\delta_{K}^2}{1-\delta_{k-1}} \right) \frac{\left\| \mathbf{x}_{T - s^{k-1}_i} \right\|_2}{\sqrt{K-k+1}}		\nonumber	\\
	&\geq & \left( 1- \delta_{L+K}-\frac{ \delta_{L+K}^2 } {(1 - \delta _{L+K})} \right) \frac{\left\| \mathbf{x}_{T - s^{k-1}_i} \right\|_2}{\sqrt{K-k+1}}	\nonumber	\\
	&=& \frac{ 1 - 2\delta_{L+K} } {1 - \delta _{L+K}} \frac{\left\| \mathbf{x}_{T - s^{k-1}_i} \right\|_2}{\sqrt{K-k+1}}.	 \label{eq:lem_b1_ex}
\end{IEEEeqnarray}
Using \eqref{eq:lem_aL_ex} and \eqref{eq:lem_b1_ex}, we obtain the sufficient condition of \eqref{eq:sc_gen_sketch} as
\begin{equation}
	\frac{ 1 - 2\delta_{L+K} } {1 - \delta _{L+K}} \frac{\left\| \mathbf{x}_{T - s^{k-1}_i} \right\|_2}{\sqrt{K-k+1}} > \frac{\delta _{L+K}} {1 - \delta_{L+K}} \frac{\left \| \mathbf{x}_{T - s^{k-1}_i} \right \|_2 }{\sqrt{L}}.
\label{eq:newtome}
\end{equation}
Rearranging \eqref{eq:newtome}, we further have 
\begin{equation}
	\delta_{L+K} < \frac{\sqrt{L}}{\sqrt{K-k+1} + 2\sqrt{L}}.	\label{eq:sc_gen_after_man}
\end{equation}
Since $\sqrt{K-k+1} < \sqrt{K}$ for $k>1$, \eqref{eq:sc_gen_after_man} holds under
$\delta_{L+K} < \frac{\sqrt{L}}{\sqrt{K} + 2\sqrt{L}}$,
which completes the proof.
\end{IEEEproof}

\subsection{Overall Sufficient Condition}

In Theorems \ref{thm:sc_init} and \ref{thm:sc_gen}, we obtained the RIP based recovery conditions guaranteeing the success of the MMP algorithm in the initial iteration $\left( \delta_{K+L} < \frac{\sqrt{L}}{\sqrt{K}+\sqrt{L}} \right)$ and non-initial iterations $\left( \delta_{K+L} < \frac{\sqrt{L}}{\sqrt{K}+2\sqrt{L}}\right)$.
Following theorem states the overall condition of MMP ensuring the accurate recovery of $K$-sparse signals.

\begin{theorem}\label{thm:sc_final}
MMP recovers $K$-sparse signal $\mathbf{x}$ from the measurements $\mathbf{y} = \mathbf{\Phi x}$ accurately if the sensing matrix satisfies the RIP with
\begin{equation}
\delta_{K+L} < \frac{\sqrt{L}}{\sqrt{K}+2\sqrt{L}}.
\label{eq:final}
\end{equation}
\end{theorem}
\begin{IEEEproof}
Since the stricter condition between two becomes the final recovery condition, it is immediate from Theorems \ref{thm:sc_init} and \ref{thm:sc_gen} that MMP accurately recovers $K$-sparse signals under \eqref{eq:final}.
\label{rev_page:13}
\end{IEEEproof}

\begin{remark}\label{cor:cond_LK}
When $L=K$, the perfect recovery condition of MMP becomes $\delta_{2K} < 0.33$.
When compared to the conditions of the CoSaMP algorithm ($\delta_{4K} < 0.1$) \cite{needell2009cosamp}, the SP algorithm ($\delta_{3K} < 0.165$) \cite{dai2009subspace}, the ROMP algorithm ($\delta_{2K} < \frac{0.03}{\sqrt{\log{2K}}}$) \cite{needell2009uniform}, and the gOMP algorithm for $N=K$ $\left( \delta_{K^2} < 0.25 \right)$ \cite{wang2012gomp}, we observe that the MMP algorithm provides more relaxed recovery condition, which in turn implies that the set of sensing matrices for which the exact recovery of the sparse signals is ensured gets larger.\footnote{Note that since these conditions are sufficient, not necessary and sufficient, direct comparison is not strictly possible.}
\end{remark}

\section{Recovery from noisy measurements} \label{sec:noisy_anal}

In this section, we investigate the RIP based condition of MMP to identify the true support set $T$ from noisy measurements $\mathbf{y} = \mathbf{\Phi}\mathbf{x}+\mathbf{v}$ where $\mathbf{v}$ is the noise vector.
In contrast to the noiseless scenario, our analysis is divided into three parts.
In the first and second parts, we consider conditions guaranteeing the success in the initial iteration ($k=1$) and non-initial iterations ($k>1$).
Same as the noiseless scenario, the success means that an index in $T$ is chosen in the iteration.
While the candidate whose magnitude of the residual is minimal (which corresponds to the output of MMP) becomes the true support in the noiseless scenario, such is not the case for the noisy scenario.
Therefore, other than two conditions we mentioned, we need an additional condition for the identification of the true support in the last minute.\footnote{Note that the accurate identification of the support $T$ cannot be translated into the perfect recovery of the original sparse signal due to the presence of the noise.}
Indeed, noting that one of candidates generated by MMP is the true support $T$ by two conditions, what we need is a condition under which the candidate whose magnitude of the residual is minimal becomes the true support.
By choosing the strictest condition among three, we obtain the condition to identify the true support in the noisy scenario.	 \label{rev_page:14}

\subsection{Success Condition in the First Iteration} \label{sub_sec:noisy_anal_init}

Recall that in the first iteration, MMP computes the correlation between $\mathbf{y}$ and $\mathbf{\phi}_i$ and then selects $L$ indices of columns having largest correlation in magnitude.
The following theorem provides a sufficient condition to ensure the success of MMP in the first iteration.
\begin{theorem}\label{thm:sc_init_n}
If all nonzero elements $x_i$ in the $K$-sparse vector $\mathbf{x}$ satisfy
\begin{equation}
|x_i| > \gamma \|\mathbf{v}\|_2
\label{eq:sc_init_n}
\end{equation}
where $\gamma=\frac{\sqrt{1+\delta_{L+K}}(\sqrt{L}+\sqrt{K})} {\sqrt{LK}-(\sqrt{LK}+K)\delta_{L+K} }$,
then among $L$ candidates at least one contains true index in the first iteration of MMP.
\end{theorem}
\begin{IEEEproof}
Before we proceed, we present a brief overview of the proof. Recalling the definition of $\alpha_L^k$ in \eqref{eq:def_aL}, it is clear that a (sufficient) condition of MMP for choosing at least one correct index in the first iteration is
\begin{equation}
\left\| \mathbf{\Phi}_T' \mathbf{y} \right\|_{\infty} > \alpha_L^1	\label{eq:init_cond_n}
\end{equation}
where $\alpha_L^1$ is the $L$-th largest correlation between $\mathbf{y}$ and $\mathbf{\Phi}_{T^C}$.
If we denote an upper bound $\alpha_L^1$ as $B_u$ (i.e., $B_u \geq \alpha_L^1$) and a lower bound $\left\| \mathbf{\Phi}_T' \mathbf{y} \right\|_{\infty}$ as $B_l$ (i.e., $\left\| \mathbf{\Phi}_T' \mathbf{y} \right\|_{\infty} \geq B_l$),
then it is clear that \eqref{eq:init_cond_n} holds true under $B_l>B_u$.
Since both $B_l$ and $B_u$ are a function of $\mathbf{x}_T$, we can obtain the success condition in the first iteration, expressed in terms of $\mathbf{x}_T$ (see \eqref{eq:sc_init_n}).

First, using the norm inequality, we have
\begin{IEEEeqnarray}{rCl}
\left\| \mathbf{\Phi}_{F_L}' \mathbf{y} \right\|_2
	& \geq & \frac{\left\| \mathbf{\Phi}_{F_L}' \mathbf{y} \right\|_1}{\sqrt{L}}	 \nonumber	\\
	& \geq & \frac{L  \alpha_L^1 }{\sqrt{L}}	= \sqrt{L}  \alpha_L^1.	\label{eq:n_init_l_2}
\end{IEEEeqnarray}
Also,
\begin{IEEEeqnarray}{rCl}
\left\| \mathbf{\Phi}_{F_L}' \mathbf{y} \right\|_2 &=& \left\| \mathbf{\Phi}_{F_L}' (\mathbf{\Phi}_T\mathbf{x}_T+\mathbf{v}) \right\|_2		\nonumber	\\
	&\leq& \left\| \mathbf{\Phi}_{F_L}' \mathbf{\Phi}_T\mathbf{x}_T \right\|_2 + \left\| \mathbf{\Phi}_{F_L}'\mathbf{v} \right\|_2	\nonumber	\\
	&\overset{ (a) }{\leq}& \delta_{L+K} \left\| \mathbf{x}_T \right\|_2 + \left\| \mathbf{\Phi}_{F_L}'\mathbf{v} \right\|_2	 \nonumber	\\
	&\leq& \delta_{L+K} \left\| \mathbf{x}_T \right\|_2 + \sqrt{1+\delta_{L}}\left\| \mathbf{v} \right\|_2		\nonumber	 \\
	&\overset{ (b) }{\leq}& \delta_{L+K} \left\| \mathbf{x}_T \right\|_2 + \sqrt{1+\delta_{L+K}}\left\| \mathbf{v} \right\|_2	 \label{eq:n_init_l_1}
\end{IEEEeqnarray}
where (a) and (b) follow from Lemma \ref{lem:rip_add} and \ref{lem:rip_mono}, respectively.
Using \eqref{eq:n_init_l_2} and \eqref{eq:n_init_l_1}, we obtain an upper bound of $\alpha_L^1$ as
\begin{equation}
 \alpha_L^1 \leq B_u = \frac{\delta_{L+K} \left\| \mathbf{x}_T \right\|_2 + \sqrt{1+\delta_{L+K}}\left\| \mathbf{v} \right\|_2}{\sqrt{L}}	\label{eq:n_init_l}.
\end{equation}

We next consider a lower bound $B_l$ of $\left\| \mathbf{\Phi}_T' \mathbf{y} \right\|_{\infty}$.
First, it is clear that
\begin{equation}
\left\| \mathbf{\Phi}_T' \mathbf{y} \right\|_{\infty} \geq \frac{\left\| \mathbf{\Phi}_T' \mathbf{y} \right\|_2 }{\sqrt{K}}.	 \label{eq:n_init_u_1_p}
\end{equation}
Furthermore,
\begin{IEEEeqnarray}{rCl}
\ifCLASSOPTIONonecolumn
\left\| \mathbf{\Phi}_T' \mathbf{y} \right\|_2 &=& \left\| \mathbf{\Phi}_T' \left(\mathbf{\Phi}_T\mathbf{x}_T+\mathbf{v} \right)\right\|_2	\nonumber		\\
\else
\lefteqn{ \left\| \mathbf{\Phi}_T' \mathbf{y} \right\|_2 = \left\| \mathbf{\Phi}_T' \left(\mathbf{\Phi}_T\mathbf{x}_T+\mathbf{v} \right)\right\|_2 }	\nonumber		\\
\fi
	&\geq& \left\| \mathbf{\Phi}_T' \mathbf{\Phi}_T\mathbf{x}_T \right \|_2 - \left\| \mathbf{\Phi}_T'\mathbf{v} \right\|_2	 \nonumber		\\
	&\geq& (1-\delta_K)\left\| \mathbf{x}_T \right \|_2 - \left\| \mathbf{\Phi}_T'\mathbf{v} \right\|_2		 \nonumber	 \\
	&\geq& (1-\delta_K)\left\| \mathbf{x}_T \right \|_2 - \sqrt{1+\delta_K}\left\| \mathbf{v} \right\|_2	 \nonumber	 \\
	&\geq& (1-\delta_{L+K})\left\| \mathbf{x}_T \right \|_2 - \sqrt{1+\delta_{L+K}} \left\| \mathbf{v} \right\|_2	 \label{eq:n_init_u_1}.
\end{IEEEeqnarray}
From \eqref{eq:n_init_u_1_p} and \eqref{eq:n_init_u_1}, we obtain a lower bound of $\left\| \mathbf{\Phi}_T' \mathbf{y} \right\|_{\infty}$ as
\begin{equation}
\left\| \mathbf{\Phi}_T' \mathbf{y} \right\|_{\infty}
	\geq B_l = \frac{(1-\delta_{L+K})\left\| \mathbf{x}_T \right \|_2 - \sqrt{1+\delta_{L+K}}\left\| \mathbf{v} \right\|_2}{\sqrt{K}}	\label{eq:n_init_u}.
\end{equation}

So far, we have obtained $B_u$ in \eqref{eq:n_init_l} and $B_l$ in \eqref{eq:n_init_u}.
Using these, 
we obtain the sufficient condition of \eqref{eq:init_cond_n} as

\ifCLASSOPTIONonecolumn
\begin{equation}
\frac{(1-\delta_{L+K})\left\| \mathbf{x}_T \right \|_2 - \sqrt{1+\delta_{L+K}}\left\| \mathbf{v} \right\|_2}{\sqrt{K}}
>
\frac{\delta_{L+K} \left\| \mathbf{x}_T \right\|_2 + \sqrt{1+\delta_{L+K}}\left\| \mathbf{v} \right\|_2}{\sqrt{L}}.
\end{equation}
\else
\begin{IEEEeqnarray}{rCl}
\frac{(1-\delta_{L+K})\left\| \mathbf{x}_T \right \|_2 - \sqrt{1+\delta_{L+K}}\left\| \mathbf{v} \right\|_2}{\sqrt{K}}	\nonumber	 \\
> \frac{\delta_{L+K} \left\| \mathbf{x}_T \right\|_2 + \sqrt{1+\delta_{L+K}}\left\| \mathbf{v} \right\|_2}{\sqrt{L}}.
\end{IEEEeqnarray}
\fi
%
After some manipulations, we have
\begin{equation}
\| \mathbf{x}_T \|_2 > \frac{\sqrt{1+\delta_{L+K}} (\sqrt{L}+\sqrt{K})}{\sqrt{L}-(\sqrt{L}+\sqrt{K})\delta_{L+K} } \|\mathbf{v}\|_2	\label{eq:n_init_t}.
\end{equation}
Since $\|\mathbf{x}_{T}\|_2^2 = \sum_{j \in T} x_j^2 \geq \left| T \right | \min_{i \in T} \left | x_i \right |^2 = K \min_{i \in T} \left | x_i \right |^2 $, \eqref{eq:n_init_t} holds if
\begin{equation}
\min_{i \in T}|x_i| > \frac{\sqrt{1+\delta_{L+K}}(\sqrt{L}+\sqrt{K})} { \sqrt{K} \left(\sqrt{L}-(\sqrt{L}+\sqrt{K})\delta_{L+K} \right)} \|\mathbf{v}\|_2,	\label{rev_eq:53}
\end{equation}
which completes the proof.
\end{IEEEproof}

\subsection{Success Condition in Non-initial Iterations} \label{sub_sec:noisy_anal_n_init}

When compared to the analysis for the initial iteration, non-initial iteration part requires an extra effort in the construction of the upper bound of $\alpha_L^K$ and the lower bound of $\beta_1^K$.
%
\begin{theorem}\label{thm:sc_gen_n}
If all the nonzero elements $x_i$ in the $K$-sparse vector $\mathbf{x}$ satisfy
\begin{equation}
|x_i| > \mu \|\mathbf{v}\|_2
\label{eq:sc_gen_n}
\end{equation}
where $\mu=\frac{\sqrt{1+\delta_{L+K}} (1-\delta_{L+K}) (\sqrt{L}+\sqrt{K})}{\sqrt{L}-(2\sqrt{L}+\sqrt{K})\delta_{L+K} }$,
then among $L$ candidates at least one contains the true index in the $k$-th iteration of MMP.
\end{theorem}

Similar to the analysis of the noiseless scenario,
key point of the proof is that $\alpha_L^k < \beta_1^k$ ensures the success in $k$-th iteration.
In the following two lemmas, we construct a lower bound of $\alpha_L^k$ and an upper bound of $\beta_1^k$, respectively.
%
\begin{lemma}\label{lem:aL_noisy}
Suppose a candidate $s^{k-1}_i$ includes indices only in $T$, then $\alpha_L^k$ satisfies
\ifCLASSOPTIONonecolumn
\begin{equation}
\alpha_L^k \leq \left( \delta _{L+K-k+1} + \frac{\delta _{L+k-1} \delta_{K}} {1 - \delta_{k-1}} \right ) \frac{\left \| \mathbf{x}_{T - s^{k-1}_i} \right \|_2 }{\sqrt{L}} + \frac{\sqrt{ 1+\delta_L } \left\| \mathbf{v} \right\|_2}{\sqrt{L}}.
\end{equation}
\else
\begin{IEEEeqnarray}{rCl}
\alpha_L^k &\leq& \left( \delta _{L+K-k+1} + \frac{\delta _{L+k-1} \delta_{K}} {1 - \delta_{k-1}} \right ) \frac{\left \| \mathbf{x}_{T - s^{k-1}_i} \right \|_2 }{\sqrt{L}}	\nonumber	\\
 & &+ \frac{\sqrt{ 1+\delta_L } \left\| \mathbf{v} \right\|_2}{\sqrt{L}}.
\end{IEEEeqnarray}
\fi
\end{lemma}
\begin{IEEEproof}
See Appendix \ref{app:aL_n}
\end{IEEEproof}

\begin{lemma}\label{lem:b1_noisy}
Suppose a candidate $s^{k-1}_i$ includes indices only in $T$, then $\beta_1^k$ satisfies
\ifCLASSOPTIONonecolumn
\begin{equation}
\beta_1^k
\geq
\left( 1- \delta_{K-k+1}-\frac{\delta_{K}^2}{1-\delta_{k-1}} \right) \frac{\left\| \mathbf{x}_{T - s^{k-1}_i} \right\|_2}{\sqrt{K-k+1}}-\frac{\sqrt{1+\delta_{K-k+1}} \left \| \mathbf{v} \right \|_2}{\sqrt{K-k+1}}.	 \label{rev_eq:56}
\end{equation}
\else
\begin{IEEEeqnarray}{rCl}
\beta_1^k
&\geq&
\left( 1- \delta_{K-k+1}-\frac{\sqrt{1+\delta_{K-k+1}} \sqrt{1+\delta_{k-1}} \delta_{K}}{1-\delta_{k-1}} \right) 	 \nonumber	 \\
 & & \cdot \frac{\left\| \mathbf{x}_{T - s^{k-1}_i} \right\|_2}{\sqrt{K-k+1}}-\frac{\sqrt{1+\delta_{K-k+1}} \left \| \mathbf{v} \right \|_2}{\sqrt{K-k+1}}.	 \label{rev_eq:56}
\end{IEEEeqnarray}
\fi
\end{lemma}
\begin{IEEEproof}
See Appendix \ref{app:b1_n}
\end{IEEEproof}

\begin{IEEEproof}[\textbf{Proof of Theorem \ref{thm:sc_gen_n}}]

Using Lemma \ref{lem:aL_noisy}, we have
\ifCLASSOPTIONonecolumn
\begin{IEEEeqnarray}{rCl}
\alpha_L^k
	&\leq & \left( \delta _{L+K - k+1} + \frac{\delta _{L+k-1} \delta_{K}} {1 - \delta_{k-1}} \right ) \frac{\left \| \mathbf{x}_{T - s^{k-1}_i} \right \|_2 }{\sqrt{L}}  + \frac{\sqrt{ 1+\delta_L } \left\| \mathbf{v} \right\|_2}{\sqrt{L}}		 \\
	&\leq & \left( \delta _{L+K} + \frac{\delta _{L+K} \delta_{L+K}} {1 - \delta_{L+K}} \right ) \frac{\left \| \mathbf{x}_{T - s^{k-1}_i} \right \|_2 }{\sqrt{L}}  + \frac{\sqrt{ 1+\delta_{L+K} } \left\| \mathbf{v} \right\|_2}{\sqrt{L}}	 \label{eq:lem_aL_noisy_ex_1}	\\
	&= & \frac{\delta _{L+K}} {1 - \delta_{L+K}} \frac{\left \| \mathbf{x}_{T - s^{k-1}_i} \right \|_2 }{\sqrt{L}}  + \frac{\sqrt{ 1+\delta_{L+K} } \left\| \mathbf{v} \right\|_2}{\sqrt{L}}	 \label{eq:lem_aL_noisy_ex}
\end{IEEEeqnarray}
\else
\begin{IEEEeqnarray}{rCl}
\alpha_L^k
	&\leq & \left( \delta _{L+K - k+1} + \frac{\delta _{L+k-1} \delta_{K}} {1 - \delta_{k-1}} \right ) \frac{\left \| \mathbf{x}_{T - s^{k-1}_i} \right \|_2 }{\sqrt{L}}	\nonumber	\\
	& & + \frac{\sqrt{ 1+\delta_L } \left\| \mathbf{v} \right\|_2}{\sqrt{L}}		\\
	&\leq & \left( \delta _{L+K} + \frac{\delta _{L+K} \delta_{L+K}} {1 - \delta_{L+K}} \right ) \frac{\left \| \mathbf{x}_{T - s^{k-1}_i} \right \|_2 }{\sqrt{L}}	\nonumber	\\
	& &+ \frac{\sqrt{ 1+\delta_{L+K} } \left\| \mathbf{v} \right\|_2}{\sqrt{L}}	 \label{eq:lem_aL_noisy_ex_1}	\\
	&= & \frac{\delta _{L+K}} {1 - \delta_{L+K}} \frac{\left \| \mathbf{x}_{T - s^{k-1}_i} \right \|_2 }{\sqrt{L}} + \frac{\sqrt{ 1+\delta_{L+K} } \left\| \mathbf{v} \right\|_2}{\sqrt{L}}	 \label{eq:lem_aL_noisy_ex}
\end{IEEEeqnarray}
\fi
where \eqref{eq:lem_aL_noisy_ex_1} follows from the monotonicity of the restricted isometry constant. 
Using Lemma \ref{lem:b1_noisy}, we also have
\begin{IEEEeqnarray}{rCl}
\beta_1^k
	&\geq & \left( 1- \delta_{K-k+1}-\frac{\delta_{K}^2}{1-\delta_{k-1}} \right) \frac{\left\| \mathbf{x}_{T - s^{k-1}_i} \right\|_2}{\sqrt{K-k+1}} \nonumber	\\
	& & -\frac{\sqrt{1+\delta_{K-k+1}} \left \| \mathbf{v} \right \|_2}{\sqrt{K-k+1}}		\\
	&\geq & \left( 1- \delta_{L+K}-\frac{ \delta_{L+K}^2 } {(1 - \delta _{L+K})} \right) \frac{\left\| \mathbf{x}_{T - s^{k-1}_i} \right\|_2}{\sqrt{K-k+1}}	 	\nonumber	\\
	& & -\frac{\sqrt{1+\delta_{L+K}} \left \| \mathbf{v} \right \|_2}{\sqrt{K-k+1}}		 \label{eq:lem_b1_noisy_ex_1}	\\
	&=& \frac{ 1 - 2\delta_{L+K} } {1 - \delta _{L+K}} \frac{\left\| \mathbf{x}_{T - s^{k-1}_i} \right\|_2}{\sqrt{K-k+1}} -\frac{\sqrt{1+\delta_{L+K}} \left \| \mathbf{v} \right \|_2}{\sqrt{K-k+1}}	 \label{eq:lem_b1_noisy_ex}
\end{IEEEeqnarray}
where \eqref{eq:lem_b1_noisy_ex_1} follows from Lemma \ref{lem:rip_mono}.

Now, using \eqref{eq:lem_aL_noisy_ex} and \eqref{eq:lem_b1_noisy_ex}, we obtain the sufficient condition of $\beta_1^k > \alpha_L^k$ as
\ifCLASSOPTIONonecolumn
\begin{equation}
	\left\| \mathbf{x}_{T - s^{k-1}_i} \right\|_2 > \frac{\sqrt{1+\delta_{L+K}} (1-\delta_{L+K}) (\sqrt{L}+\sqrt{K-k+1})}{\sqrt{L}-(2\sqrt{L}+\sqrt{K-k+1})\delta_{L+K} } \left\| \mathbf{v} \right\|_2. \label{eq:sc_gen_n_after_man}
\end{equation}
\else
\begin{IEEEeqnarray}{rCl}
	\lefteqn{ \left\| \mathbf{x}_{T - s^{k-1}_i} \right\|_2 > } \nonumber	\\
	 & & \frac{\sqrt{1+\delta_{L+K}} (1-\delta_{L+K}) (\sqrt{L}+\sqrt{K-k+1})}{\sqrt{L}-(2\sqrt{L}+\sqrt{K-k+1})\delta_{L+K} } \left\| \mathbf{v} \right\|_2. \label{eq:sc_gen_n_after_man}
\end{IEEEeqnarray}
\fi
%
In the non-initial iterations ($2 \leq k \leq K$), $\sqrt{K-k+1} < \sqrt{K}$ and $$\|\mathbf{x}_{T-s^{k-1}_i}\|_2^2 = \sum_{j \in T-s^{k-1}_i} \left | x_j \right |^2 \geq \left | T-s^{k-1}_i \right | \min_{i \in T} \left | x_i \right |^2 \geq \left | x_i \right |^2,$$ so that \eqref{eq:sc_gen_n_after_man} holds if
\begin{equation}
	\min_{i \in T} \left | x_i \right | > \frac{\sqrt{1+\delta_{L+K}} (1-\delta_{L+K}) (\sqrt{L}+\sqrt{K})}{\sqrt{L}-(2\sqrt{L}+\sqrt{K})\delta_{L+K} } \left\| \mathbf{v} \right\|_2	 \label{rev_eq:64}
\end{equation}
which is the desired result.
\end{IEEEproof}

\subsection{Condition in the Final Stage}

As mentioned, in the noiseless scenario, the candidate whose magnitude of the residual is minimal becomes the true support $T$.
In other words, if $s^* = \arg \min_s \left \| \mathbf{r}_s \right \|^2_2$, then $\left \| \mathbf{r}_{s^*} \right \|^2_2=0$ and also $s^*=T$.
Since this is not true for the noisy scenario, an additional condition ensuring the selection of true support is required.
The following theorem provides a condition under which the output of MMP (candidate whose magnitude of the residual is minimal) becomes the true support.
%

\begin{theorem}\label{thm:resi_small}
If all the nonzero coefficients $x_i$ in the $K$-sparse signal vector $\mathbf{x}$ satisfy
\begin{equation}
|x_i| \geq \lambda \|\mathbf{v}\|_2
\label{eq:n_cond}
\end{equation}
where $\lambda=\sqrt{\frac{2 (1-\delta_K)^2}{(1-\delta_K)^3-
		(1+\delta_K)\delta_{2K}^2}}$, then
\begin{equation}
\| \mathbf{r}_T \| \leq \min_{\Gamma \in \Omega^K} \| \mathbf{r}_{\Gamma} \|	\label{eq:n_cond_str}
\end{equation}
 where $\Omega^K$ is the set of all possible combinations of $K$ columns in $\mathbf{\Phi}$.
\end{theorem}

\begin{IEEEproof}
First, one can observe that an upper bound of $\left \| \mathbf{r}_T \right \|_2^2$ is
\begin{IEEEeqnarray}{rCl}
\| \mathbf{r}_T \|_2^2 &=& \| \mathbf{P}_T^\bot \mathbf{y} \|_2^2	\nonumber \\
	&=& \| \mathbf{P}_T^\bot \left( \mathbf{\Phi}_T\mathbf{x}_T + \mathbf{v} \right) \|_2^2	\nonumber \\
	&=& \| \mathbf{P}_T^\bot \mathbf{\Phi}_T\mathbf{x}_T + \mathbf{P}_T^\bot \mathbf{v} \|_2^2	\nonumber \\
	&=& \| (\mathbf{\Phi}_T\mathbf{x}_T - \mathbf{P}_T \mathbf{\Phi}_T\mathbf{x}_T) + \mathbf{P}_T^\bot \mathbf{v} \|_2^2 \nonumber \\
	&=& \| \mathbf{\Phi}_T\mathbf{x}_T - \mathbf{\Phi}_T (\mathbf{\Phi}_T'\mathbf{\Phi}_T)^{-1} \mathbf{\Phi}_T' \mathbf{\Phi}_T\mathbf{x}_T + \mathbf{P}_T^\bot \mathbf{v} \|_2^2	\nonumber \\
	&=& \|\mathbf{P}_T^{\bot}\mathbf{v}\|_2^2	\nonumber \\
	&\leq& \|\mathbf{v}\|_2^2	\label{eq:rt}.
\end{IEEEeqnarray}
Next, a lower bound of $\left \| \mathbf{r}_\Gamma \right \|_2^2$ is (see Appendix \ref{app:lb_resi})
\ifCLASSOPTIONonecolumn
\begin{equation}
\left( (1-\delta_{|T-\Gamma|})-
		\frac{(1+\delta_{|\Gamma|})\delta_{|\Gamma|+|T-\Gamma|}^2}{(1-\delta_{|\Gamma|})^2} \right) \|\mathbf{x}_{T-\Gamma}\|_2^2 - \|\mathbf{v}\|_2^2
\leq
\| \mathbf{r}_{\Gamma} \|_2^2	\label{eq:rl}.
\end{equation}
\else
\begin{IEEEeqnarray}{rCl}
\left( (1-\delta_{|T-\Gamma|})-
		\frac{(1+\delta_{|\Gamma|})\delta_{|\Gamma|+|T-\Gamma|}^2}{(1-\delta_{|\Gamma|})^2} \right) \|\mathbf{x}_{T-\Gamma}\|_2^2 - \|\mathbf{v}\|_2^2	\nonumber	\\
\leq
\| \mathbf{r}_{\Gamma} \|_2^2.	\IEEEeqnarraynumspace	\label{eq:rl}
\end{IEEEeqnarray}
\fi
%
From \eqref{eq:rt} and \eqref{eq:rl}, it is clear that \eqref{eq:n_cond_str} is achieved if
\begin{equation}
\|\mathbf{x}_{T-\Gamma}\|_2^2 \geq \frac{2\|\mathbf{v}\|_2^2}{(1-\delta_{|T-\Gamma|})-
		\frac{(1+\delta_{|\Gamma|})\delta_{|\Gamma|+|T-\Gamma|}^2}{(1-\delta_{|\Gamma|})^2}}.	 \label{eq:n_cond_pn}
\end{equation}
Further, noting that $|\Gamma|=K$, $|T-\Gamma| \leq K$, and $|\Gamma|+|T-\Gamma| \leq 2K$, one can see that \eqref{eq:n_cond_pn} is guaranteed under
\begin{equation}
\|\mathbf{x}_{T-\Gamma}\|_2^2 \geq
		\frac{2 (1-\delta_K)^2 \|\mathbf{v}\|_2^2}{(1-\delta_K)^3- (1+\delta_K)\delta_{2K}^2}.	 \label{eq:n_cond_man}
\end{equation}
Finally, since $\|\mathbf{x}_{T-\Gamma}\|_2^2 = \sum_{j \in T-\Gamma} x_j^2 \geq \left| T-\Gamma \right | \min_{i \in T} \left | x_i \right |^2$, \eqref{eq:n_cond_man} holds true under
\begin{equation}
	\min_{i \in T} \left | x_i \right |_2^2 \geq \frac{2 (1-\delta_K)^2 \|\mathbf{v}\|_2^2}{(1-\delta_K)^3- (1+\delta_K)\delta_{2K}^2}
\end{equation}
which completes the proof.
\end{IEEEproof}

\subsection{Overall Sufficient Condition}

Thus far, we investigated conditions guaranteeing the success of the MMP algorithm in the initial iteration (Theorem \ref{thm:sc_init_n}), non-initial iterations (Theorem \ref{thm:sc_gen_n}), and also the condition that the candidate whose magnitude of the residual is minimal corresponds to the true support (Theorem \ref{thm:resi_small}).
Since one of candidates should be the true support by Theorem \ref{thm:sc_init_n} and \ref{thm:sc_gen_n} and the candidate with the minimal residual corresponds to the true support by Theorem \ref{thm:resi_small}, one can conclude that MMP outputs the true support under the strictest condition among three.
%

\begin{theorem}\label{thm:fin_res_n}
If all the nonzero coefficients $x_i$ in the sparse signal vector $\mathbf{x}$ satisfy
\begin{equation}
|x_i| \geq \zeta \| \mathbf{v} \|_2
\label{eq:maxxx}
\end{equation}
where $\zeta = \max \left ( \gamma, \mu, \lambda \right )$ and $\gamma=\frac{\sqrt{1+\delta_{L+K}}(\sqrt{L}+\sqrt{K})}{ \sqrt{LK}-(\sqrt{LK}+K)\delta_{L+K}}$, $\mu=\frac{\sqrt{1+\delta_{L+K}} (1-\delta_{L+K}) (\sqrt{L}+\sqrt{K})}{\sqrt{L}-(2\sqrt{L}+\sqrt{K})\delta_{L+K} }$, and
$\lambda=\sqrt{\frac{2 (1-\delta_K)^2}{(1-\delta_K)^3-(1+\delta_K)\delta_{2K}^2}}$, then MMP chooses the true supports $T$.
\end{theorem}
\begin{IEEEproof}
Immediate from \eqref{eq:sc_init_n}, \eqref{eq:sc_gen_n}, and \eqref{eq:n_cond}.
\end{IEEEproof}

\vspace{0.1cm}

\begin{remark}\label{rmk:ls_mmse}
When the condition in \eqref{eq:maxxx} is satisfied, which is true for high SNR regime, the true support is identified and hence all non-support elements and columns associated with these can be removed from the system model.
%
In doing so, one can obtain the overdetermined system model
\begin{equation}
\mathbf{y} = \mathbf{\Phi}_T \mathbf{x}_T + \mathbf{v}	.	\label{eq:rm_sysmodel}
\end{equation}
Interestingly, the final output of MMP is equivalent to that of the Oracle LS estimator
\begin{equation}
\hat{\mathbf{x}} = \mathbf{\Phi}^{\dagger}_T \mathbf{y}=\mathbf{x}_T + \mathbf{\Phi}^{\dagger}_T \mathbf{v}
\end{equation}
and the resulting estimation error becomes $J(\hat{\mathbf{x}})=\left \| \mathbf{x}_T - \hat{\mathbf{x}} \right \|^2 = \| \mathbf{\Phi}_T^{\dagger} \mathbf{v} \|^2$.
Also, when the {\it a prior} information on signal and noise is available, one might slightly modify the final stage and obtain the linear minimum mean square error (LMMSE) estimate. For example, if the signal and noise are uncorrelated and their correlations are $\sigma_{\mathbf{x}}^2$ and $\sigma_{\mathbf{v}}^2$, respectively, then
\begin{equation}
\hat{\mathbf{x}} = \sigma_{\mathbf{x}}^2\mathbf{\Phi}^{'}_T( \sigma_{\mathbf{x}}^2 \mathbf{\Phi}_T \mathbf{\Phi}^{'}_T + \sigma_{\mathbf{v}}^2 \mathbf{I})^{-1} \mathbf{y}.
\end{equation}
%

In short, under \eqref{eq:maxxx}, MMP fully identifies the location of nonzero elements, converting the underdetermined system into the overdetermined system in \eqref{eq:rm_sysmodel}.
As a final remark, we note that the condition \eqref{eq:maxxx} is sufficient for the stable reconstruction of $K$-sparse signals in the noisy scenario.
\end{remark}

\begin{remark}	
Under \eqref{eq:maxxx}, the output of MMP satisfies
\begin{equation}
\left \| \mathbf{x} - \hat{\mathbf{x}} \right \|_2 \leq \tau \| \mathbf{v} \|_2
\end{equation}
where $\tau=\frac{1}{\sqrt{1 - \delta_{K}}}$.
\end{remark}
\begin{IEEEproof}
Since $s^*=T$ under \eqref{eq:maxxx}, $\left \| \mathbf{x} - \hat{\mathbf{x}} \right \|_2$ equals $\left \| \mathbf{x}_T - \hat{\mathbf{x}}_{s^*} \right \|_2$ by ignoring non-support elements.
Furthermore, we have
\begin{IEEEeqnarray}{rCl}
\left \| \mathbf{x}_T - \hat{\mathbf{x}}_{s^*} \right \|_2 &\leq& \frac{\left \| \mathbf{\Phi}_T \left ( \mathbf{x}_T - \hat{\mathbf{x}}_{s^*} \right ) \right \|_2}{\sqrt{1 - \delta_{\left | T \right |}}}	 \label{eq:stb_init} \nonumber \\
	&=& \lefteqn{ \frac{\left \| \mathbf{\Phi}_T \mathbf{x}_T - \mathbf{\Phi}_T \mathbf{\Phi}^{\dagger}_{s^*} \mathbf{y} \right \|_2}{\sqrt{1 - \delta_{K}}} } \nonumber \\
	&=& \frac{\left \| \mathbf{\Phi}_T \mathbf{x}_T - \mathbf{\Phi}_T \mathbf{\Phi}^{\dagger}_{s^*} \left ( \mathbf{\Phi}_T \mathbf{x}_T + \mathbf{v} \right ) \right \|_2}{\sqrt{1 - \delta_{K}}} \nonumber	\\
	&=& \frac{\left \| \mathbf{\Phi}_T \mathbf{x}_T - \mathbf{\Phi}_T \mathbf{\Phi}^{\dagger}_{s^*} \mathbf{\Phi}_T \mathbf{x}_T - \mathbf{\Phi}_T \mathbf{\Phi}^{\dagger}_{s^*} \mathbf{v} \right \|_2}{\sqrt{1 - \delta_{K}}} \nonumber	 \\
	&=& \frac{\left \| \mathbf{\Phi}_T \mathbf{x}_T - \mathbf{\Phi}_T \mathbf{x}_T - \mathbf{P}_T \mathbf{v} \right \|_2}{\sqrt{1 - \delta_{K}}}	\nonumber \\
	&=& \frac{\left \| \mathbf{P}_T \mathbf{v} \right \|_2}{\sqrt{1 - \delta_{K}}}	\nonumber \\
	&\leq& \frac{\left \| \mathbf{v} \right \|_2}{\sqrt{1 - \delta_{K}}}, \nonumber
\end{IEEEeqnarray}
which is the desired result.
\end{IEEEproof}

\vspace{0.5cm}
\section{Depth-First MMP}\label{sec:complxt}
%
The MMP algorithm described in the previous subsection can be classified as a breadth first (BF) search algorithm that performs parallel investigation of candidates.
Although a large number of candidates is merged in the middle of the search, complexity overhead is still burdensome for high-dimensional systems and also complexity varies among different realization of $\mathbf{\Phi}$ and $\mathbf{x}$.
In this section, we propose a simple modification of the MMP algorithm to control its computational complexity.
In the proposed approach, referred to as MMP-DF, search is performed sequentially and finished if the magnitude of the residual satisfies a suitably chosen termination condition (e.g., $\| \mathbf{r}_{\ell} \|^2 = 0$ in the noiseless scenario) or the number of candidates reaches the predefined maximum value $N_{\max}$.
Owing to the serialized search and the limitation on the number of candidates being examined, the computational complexity of MMP can be strictly controlled.
For example, MMP-DF returns to OMP for $N_{\max} = 1$ and the output of MMP-DF will be equivalent to that of MMP-BF for $N_{\max} = L^K$.
%

\begin{table}[t]
\caption{Relationship between the search order and iteration order ($L=2$, $K=4$)}
\label{tab:alg2}
	\begin{center}
		\begin{tabular}{ c | c }
		\hline
		Search order $\ell$ & Layer order $\left ( c_1, c_2, c_3, c_4 \right )$ of $s^K_{\ell}$\\  \hline
		1 & $\left ( 1, 1, 1, 1 \right )$	\\  \hline
		2 & $\left ( 2, 1, 1, 1 \right )$	\\  \hline
		3 & $\left ( 1, 2, 1, 1 \right )$	\\  \hline
		4 & $\left ( 2, 2, 1, 1 \right )$	\\  \hline
		5 & $\left ( 1, 1, 2, 1 \right )$	\\  \hline
		6 & $\left ( 2, 1, 2, 1 \right )$	\\  \hline
		$\vdots$ & $\vdots$	\\  \hline
		16 & $\left ( 2, 2, 2, 2 \right )$	\\  \hline
		\end{tabular}
	\end{center}
\end{table}
Since the search operation is performed in a serialized manner, a question that naturally follows is how to set the search order of candidates to find out the solution as early as possible.
%
%
Clearly, a search strategy should offer a lower computational cost than MMP-BF, yet ensures a high chance of identifying the true support even when $N_{\max}$ is small.
Intuitively, a search mechanism should avoid exploring less promising paths and give priority to the promising paths.
To meet this goal, we propose a search strategy so called {\it modulo strategy}.
The main purpose of the modulo strategy is, while putting the search priority to promising candidates, to avoid the local investigation of candidates by diversifying search paths from the top layer. Specifically, the relationship between $\ell$-th candidate $s^K_{\ell}$ (i.e., the candidate being searched at order $\ell$) and the layer order $c_k$ of the modulo strategy is defined as
%
\begin{equation}
	\ell = 1 + \sum_{k=1}^{K} {\left ( c_k - 1 \right )L^{k-1}}.
	\label{eq:dir_idx}
\end{equation}
%
By order we mean the order of columns based on the magnitude of correlation between the column $\phi_i$ and residual $\mathbf{r}^{k}$.
Notice that there exists one-to-one correspondence between the candidate order $\ell$ and the set of layer (iteration) orders $(c_1, \ldots, c_K)$.\footnote{One can think of $(c_1 -1, ~\ldots, ~ c_k -1)$ as an unsigned $L$'s complement representation of the number $\ell-1$}
For example, if $L=2$ and $K=4$, then the set of layer orders for the first candidate $s^K_{1}$ is $(c_1 , c_2 , c_3, c_4) = (1, 1, 1, 1)$ so that $s^K_{1}$ traces the path corresponding to the best index (index with largest correlation in magnitude) for all iterations (see Fig. \ref{fig:dfs_search}).
Hence, $s^K_{1}$ will be equivalent to the path of the OMP algorithm.
%
%
Whereas, for the second candidate $s^K_2$, the set of layer orders is $(c_1, c_2, c_3, c_4) = (2, 1, 1, 1)$ so that the second best index is chosen in the first iteration and the best index is chosen for the rest.
%

\begin{figure*}
	\begin{center}
		\includegraphics[width=170mm]{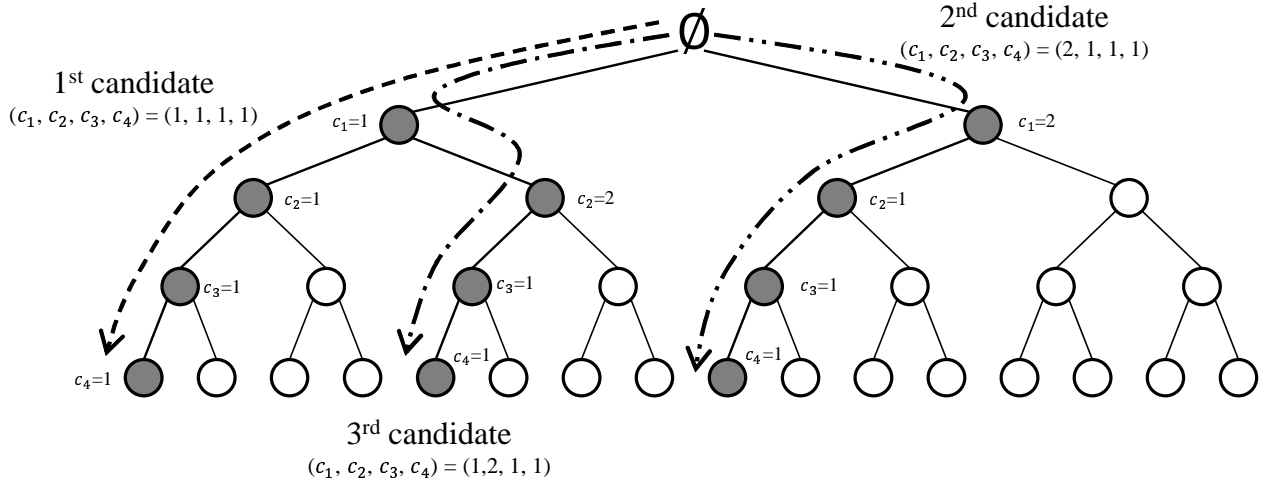}
		\caption{Illustration of MMP-DF operation ($L=2$ and $K=4$).}
		\label{fig:dfs_search}
	\end{center}
\end{figure*}

\begin{table*}
\caption{The MMP-DF algorithm}
\label{tab:alg_df}
\centering
\normalsize
\begin{tabular}{l r}
\hline
\hline
\multicolumn{2}{l}{\textbf{Input:}	}\\
\multicolumn{2}{l}{	\hspace{4mm}Measurement $\mathbf{y}$, sensing matrix $\mathbf{\Phi}$, sparsity $K$, number of expansion $L$,}\\
\multicolumn{2}{l}{	\hspace{4mm} stop threshold $\epsilon$, max number of search candidate $\ell_{\max}$		 }\\
\multicolumn{2}{l}{\textbf{Output:}	}\\
\multicolumn{2}{l}{	\hspace{4mm}Estimated signal $\hat{\mathbf{x}}$	}\\
\multicolumn{2}{l}{\textbf{Initialization:}	}\\
\multicolumn{2}{l}{	\hspace{4mm}$\ell:=0$ (candidate order), $\rho := \infty$ (min. magnitude of residual)	}\\
\hline
\multicolumn{2}{l}{{\bf while} {$\ell < \ell_{\max}$ and $\epsilon < \rho$}  {\bf do}										 }\\
\multicolumn{2}{l}{	\hspace{4mm}$\ell :=\ell+1$					}\\
\multicolumn{2}{l}{	\hspace{4mm}$\mathbf{r}^0 := \mathbf{y}$				}\\
					\hspace{4mm}$\left[c_1, \ldots ,c_K \right ] := \text{compute\_ck}( \ell, L )$	& ({\it compute layer order})\\
					\hspace{4mm}{\bf for} $k=1$ {\bf to} $K$ {\bf do}	& ({\it investigate $\ell$-th candidate})					 \\
					\hspace{8mm}	$ \tilde{\pi} := \arg \mathop {\max} \limits_{ \left | \pi \right | = L} { \| ( \mathbf{\Phi}' \mathbf{r}^{k-1} )_{\pi} \|_2^2 } $	& ({\it choose $L$ best indices})\\
					\hspace{8mm}	$s_{\ell}^k := s_{\ell}^{k-1} \cup \left \{ \tilde{\pi}_{c_k} \right \}$		& ({\it construct a path in $k$-th layer}) \\
					\hspace{8mm}	$\hat{\mathbf{x}}^k := \mathbf{\Phi}^{\dagger}_{s_{\ell}^k} \mathbf{y}$	 & ({\it estimate $\hat{\mathbf{x}}^k$ in $k$-th layer}) \\
					\hspace{8mm}	$\mathbf{r}^k := \mathbf{y}-\mathbf{\Phi}_{s_{\ell}^k} \hat{\mathbf{x}}^k$	 & ({\it update residual}) \\
\multicolumn{2}{l}{	\hspace{4mm}{\bf end for}																		 }\\
					\hspace{4mm}{\bf if} {$| \mathbf{r}^K | < \rho $} {\bf then}	& ({\it update the smallest residual})	 \\
\multicolumn{2}{l}{	\hspace{8mm}	$\rho := | \mathbf{r}^K |$														 }\\
\multicolumn{2}{l}{	\hspace{8mm}	$\hat{\mathbf{x}}^* := \hat{\mathbf{x}}^K$										 }\\
\multicolumn{2}{l}{	\hspace{4mm}{\bf end if}																		 }\\
\multicolumn{2}{l}{{\bf end while}																					 }\\
\multicolumn{2}{l}{{\bf return} $\hat{\mathbf{x}}^*$																 }\\
\hline
\multicolumn{2}{l}{\textbf{function} compute\_ck ($\ell$, $L$)	}	\\
\multicolumn{2}{l}{\hspace{4mm}$temp := \ell-1$	}	\\
\multicolumn{2}{l}{\hspace{4mm}{\bf for} $k=1$ {\bf to} $K$ {\bf do}	}	\\
\multicolumn{2}{l}{\hspace{8mm}		$c_k :=  \mod (temp, L)+1$	}	\\
\multicolumn{2}{l}{\hspace{8mm}		$temp :=  \text{floor} (temp / L)$	}	\\
\multicolumn{2}{l}{\hspace{4mm}{\bf end for}	}		\\
\multicolumn{2}{l}{{\bf return} $\left[c_1, \ldots ,c_K \right ]$																 }\\
\multicolumn{2}{l}{\textbf{end function}}	\\
\hline
\end{tabular}
\end{table*}

Table \ref{tab:alg2} summarizes the mapping between the search order and the set of layer orders for $L=2$ and $K=4$.
When the candidate order $\ell$ is given, the set of layer orders $(c_1, \cdots, c_K)$ are determined by \eqref{eq:dir_idx}
and MMP-DF traces the path using these layer orders. Specifically, in the first layer, index corresponding to the $c_1$-th largest correlation in magnitude is chosen. After updating the residual, in the second layer, index corresponding to the $c_2$-th largest correlation in magnitude is added to the path. This step is repeated until the index in the last layer is chosen. Once the full-blown candidate $s_{\ell}^K$ is constructed, MMP-DF searches the next candidate $s_{\ell+1}^K$.
After finding $N_{\max}$ candidates, a candidate whose magnitude of the residual is minimum is chosen as the final output. The operation of MMP-DF is summarized in Table \ref{tab:alg_df}.

Although the modulo strategy is a bit heuristic in nature, we observe from numerical results in Section \ref{sec:sim} that MMP-DF performs comparable to MMP-BF.
Also, as seen in Fig. \ref{fig:2k_comp}, the number of candidates of MMP-DF is much smaller than that of MMP-BF, resulting in significant savings in the computational cost.
%

\section{Numerical Results}	\label{sec:sim}

\begin{figure}[t]
\begin{center}
\ifCLASSOPTIONonecolumn
	\includegraphics[width=140mm]{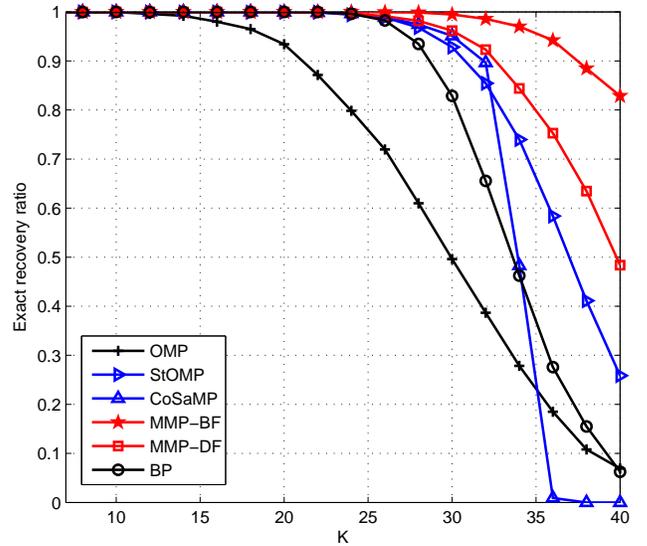}
\else
    \includegraphics[width=95mm]{gauss_err}
\fi
\caption{ERR performance of recovery algorithms as a function of sparsity $K$.} \label{fig:GaussERR}
\end{center}
\end{figure}

\begin{figure}[t]
\begin{center}
\ifCLASSOPTIONonecolumn
	\includegraphics[width=140mm]{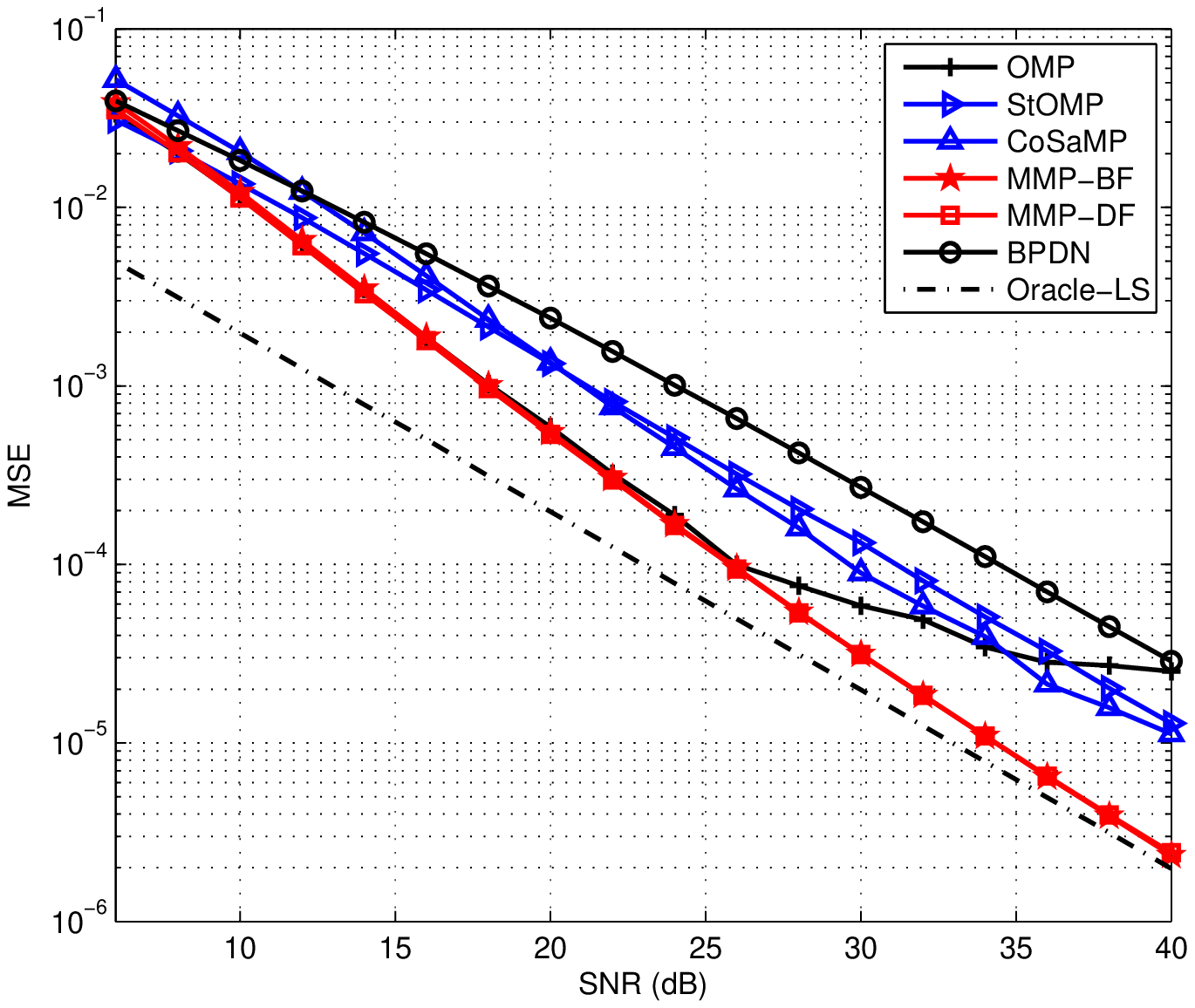}
\else
	\includegraphics[width=95mm]{gauss_sp20}
\fi
\caption{MSE performance of recovery algorithms as a function of SNR (K=20).}
\label{fig:GAUSS_SP20}
\end{center}
\end{figure}

\begin{figure}[t]
\begin{center}
\ifCLASSOPTIONonecolumn
	\includegraphics[width=140mm]{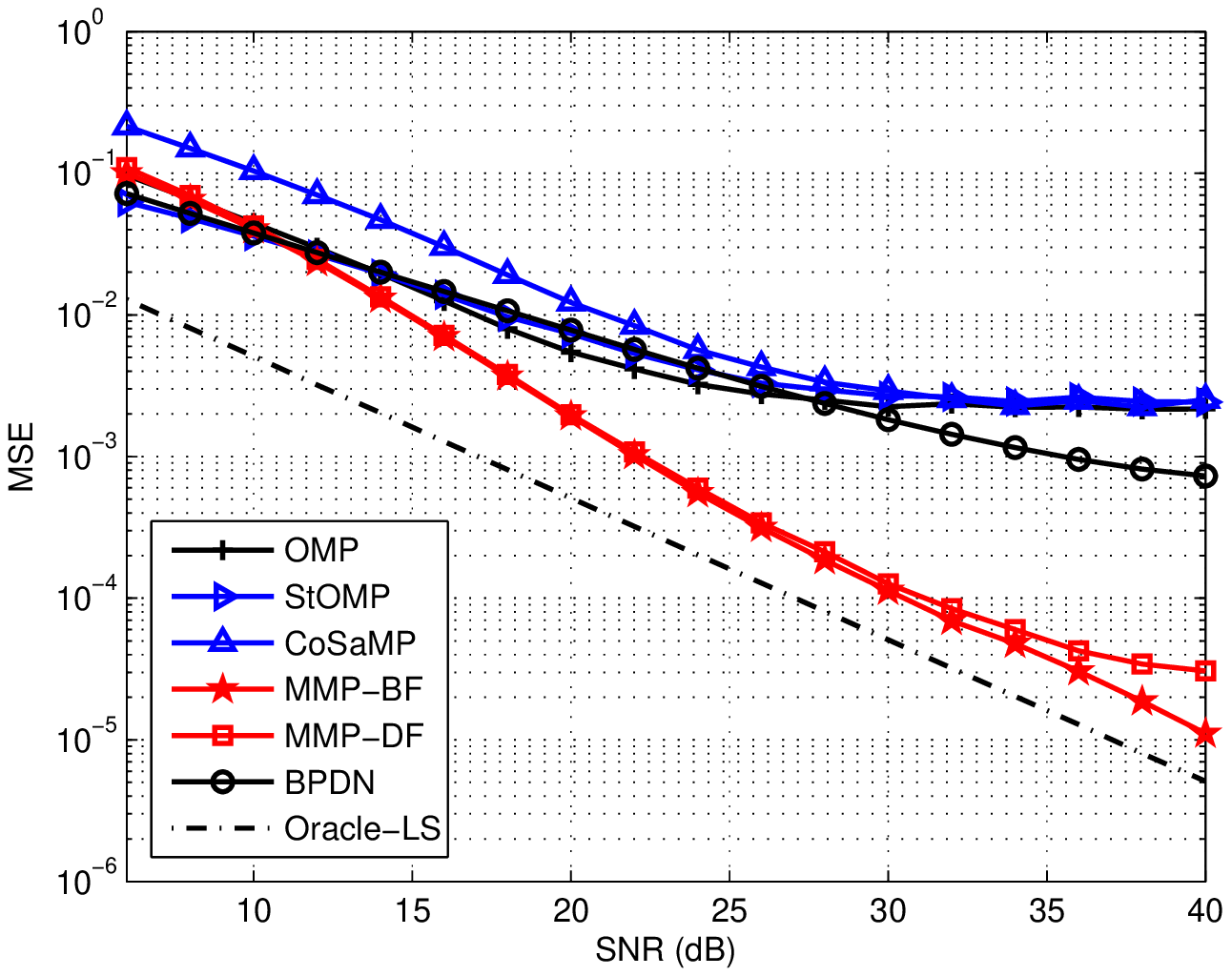}
\else
	\includegraphics[width=95mm]{gauss_sp30}
\fi
\caption{MSE performance of recovery algorithms as a function of SNR (K=30).}
\label{fig:GAUSS_SP30}
\end{center}
\end{figure}

\begin{figure}[t]
\begin{center}
	\subfigure[Miss detection]{\includegraphics[width=80mm]{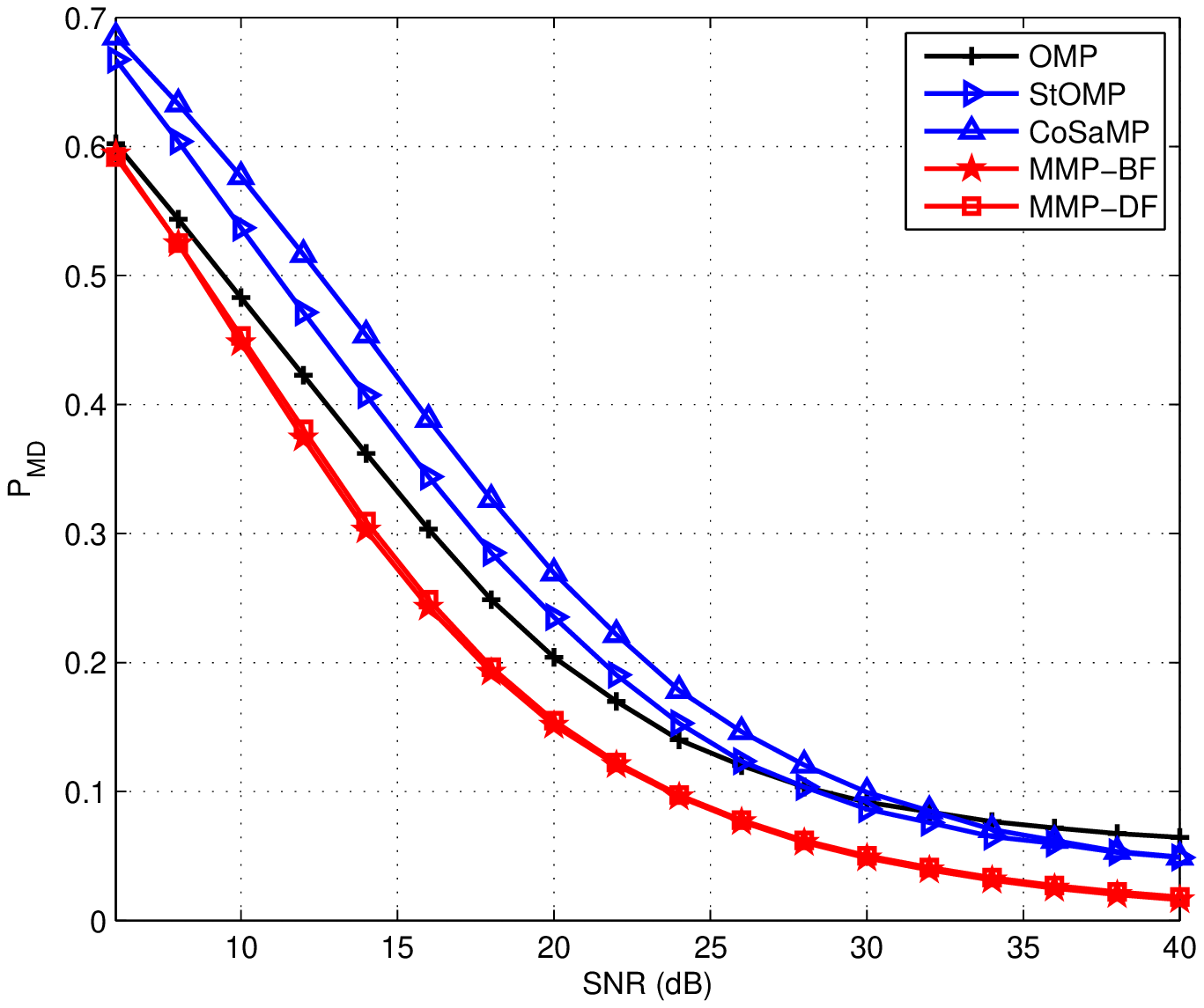} \label{fig:MD_SP30}}
	\subfigure[False alarm]{\includegraphics[width=80mm]{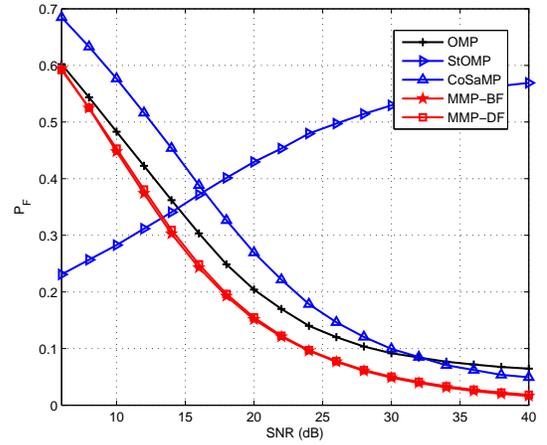} \label{fig:FA_SP30}}
\caption{Miss detection and false alarm ratio of greedy algorithms ($K=30$).}
\label{fig:MDFA_SP30}
\end{center}
\end{figure}

\begin{figure}[t]
\begin{center}
\ifCLASSOPTIONonecolumn
	\includegraphics[width=140mm]{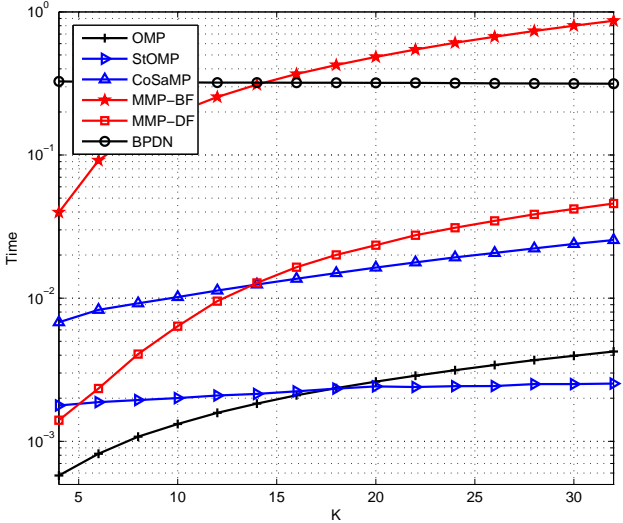}
\else
	\includegraphics[width=95mm]{gauss_time}
\fi
\caption{Running time as a function of sparsity $K$.}
\label{fig:GaussTime}
\end{center}
\end{figure}

In this section, we evaluate the performance of recovery algorithms including MMP through numerical simulations.
In our simulation, we use a random matrix $\mathbf{\Phi}$ of size $100 \times 256$ whose entries are chosen independently from Gaussian distribution $\mathcal{N}(0, 1/m)$.
We generate $K$-sparse signal vector $\mathbf{x}$ whose nonzero locations are chosen at random and their values are drawn from standard Gaussian distribution $\mathcal{N}(0, 1)$.
As a measure for the recovery performance in the noiseless scenario, \textit{exact recovery ratio} (ERR) is considered.
In the ERR experiment, accuracy of the reconstruction algorithms can be evaluated by comparing the maximal sparsity level of the underlying signals beyond which the perfect recovery is not guaranteed (this point is often called {\em critical sparsity}).
Since the perfect recovery of original sparse signals is not possible in the noisy scenario, we use the \textit{mean squared error} (MSE) as a metric to evaluate the performance:
\begin{equation}
\mbox{MSE} = \frac{1}{N} \sum_{n=1}^{N} \|\mathbf{x}[n] - \hat{\mathbf{x}}[n] \|_2^2
\end{equation}
where $\hat{\mathbf{x}}[n]$ is the estimate of the original $K$-sparse vector $\mathbf{x}[n]$ at instance $n$.
In obtaining the performance for each algorithm, we perform at least $10,000$ independent trials. In our simulations, following recovery algorithms are compared:
\begin{enumerate}
\item OMP algorithm.
\item BP algorithm: BP is used for the noiseless scenario and the basis pursuit denoising (BPDN) is used for the noisy scenario.
\item StOMP algorithm: we use false alarm rate control strategy because it performs better than the false discovery rate control strategy.
\item CoSaMP algorithm: we set the maximal iteration number to $50$ to avoid repeated iterations.
\item MMP-BF: we use $L=6$ and also set the maximal number of candidates to $50$ in each iteration.
\item MMP-DF: we set $N_{\max} = 50$.
\end{enumerate}

In Fig. \ref{fig:GaussERR}, we provide the ERR performance as a function of the sparsity level $K$.
We observe that the critical sparsities of the proposed MMP-BF and MMP-DF algorithms are larger than those of conventional recovery algorithms.
In particular, we see that MMP-BF is far better than conventional algorithms, both in critical sparsity as well as overall recovery behavior.

In Fig. \ref{fig:GAUSS_SP20}, we plot the MSE performance of sparse signals ($K=20$) in the noisy scenario as a function of the SNR where the SNR (in dB) is defined as $\mbox{SNR} = 10 \log_{10} \frac{ \| \mathbf{\Phi x} \|^2 }{ \| \mathbf{v} \|^2 }$.
In this case, the system model is expressed as $\mathbf{y} = \mathbf{\Phi} \mathbf{x} + \mathbf{v}$ where $\mathbf{v}$ is the noise vector whose elements are generated from Gaussian $\mathcal{N}(0, 10^{-\frac{SNR}{10}})$.
Generally, we observe that the MSE performance of the recovery algorithms improves with the SNR.
While the performance improvement of conventional greedy algorithms diminishes as the SNR increases, the performance of MMP improves with the SNR and performs close to the Oracle-LS estimator (see Remark \ref{rmk:ls_mmse}).

In Fig. \ref{fig:GAUSS_SP30}, 
we plot the MSE performance for $K = 30$ sparse signals.
Although the overall trend is somewhat similar to the result of $K = 20$, we can observe that the performance gain of MMP over existing sparse recovery algorithms is more pronounced.
In fact, one can indirectly infer the superiority of MMP by considering the results of Fig. \ref{fig:GaussERR} as the performance results for a very high SNR regime.
Indeed, as supported in Fig. \ref{fig:MD_SP30} and \ref{fig:FA_SP30}, both MMP-BF and MMP-DF have better (smaller) miss detection rate $P_{md}$ and false alarm rate $P_f$, which exemplifies the effectiveness of MMP for the noisy scenario.\footnote{We use $P_{md} = \frac{\text{Number of missed nonzero elementsindices}}{\text{Number of nonzero}}$ and $P_f = \frac{\text{Number of detected zero indices}}{\text{Number of nonzero}}$.}

Fig. \ref{fig:GaussTime} shows the running time of each recovery algorithm as a function of $K$.
The running time is measured using the MATLAB program on a personal computer under Intel Core i$5$ processor and Microsoft Windows $7$ environment.
Overall, we observe that MMP-BF has the highest running time and OMP and StOMP have the lowest running time among algorithms under test.
Due to the strict control of the number of candidates, MMP-DF achieves more than two order of magnitude reduction in complexity over MMP-BF.
When compared to other greedy recovery algorithms, we see that MMP-DF exhibits slightly higher but comparable complexity for small $K$.

\section{Conclusion and Discussion}	\label{sec:con}

\begin{figure}[t]
\begin{center}
\ifCLASSOPTIONonecolumn
	\includegraphics[width=140mm]{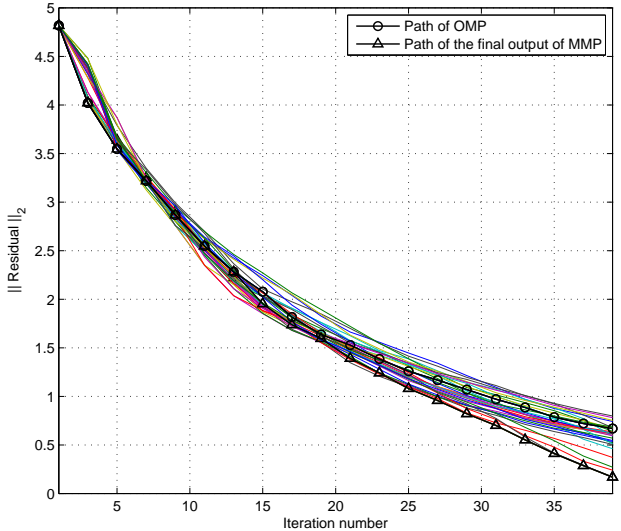}
\else
	\includegraphics[width=95mm]{candidates}
\fi
\caption{Snapshot of the magnitude of residual of candidates in MMP as a function of iteration number.}
\label{fig:Candidates}
\end{center}
\end{figure}

\subsection{Summary}
In this paper, we have taken the first step towards the exploration of multiple candidates in the reconstruction of sparse signals.
While large body of existing greedy algorithms exploits multiple indices to overcome the drawback of OMP, the proposed MMP algorithm examines multiple promising candidates with the help of greedy search.
Owing to the investigation of multiple full-blown candidates instead of partial ones, MMP improves the chance of selecting true support substantially.
In the empirical results as well as the RIP based performance guarantee, we could observe that MMP is very effective in both noisy and noiseless scenarios.

\subsection{Direction to Further Reduce Computational Overhead}
While the performance of MMP improves as the dimension of the system to be solved increases, it brings out an issue of complexity.
We already discussed in Section IV that one can strictly bound the computational complexity of MMP by serializing
the search operation and imposing limitation on the maximal number of candidates to be investigated.
As a result, we observed in Section VI that computational overhead of the modified MMP (MMP-DF) is comparable to existing greedy recovery algorithms when the sparsity level of the signal is high. 
When the signal is not very sparse, however, MMP may not be computationally appealing since most of time the number of candidates of MMP will be the same as to the predefined maximal number.

In fact, in this regime, more aggressive tree pruning strategy is required to alleviate the computational burden of MMP.
For example, in the probabilistic pruning algorithm popularly used in ML detection \cite{bshim2008sphere}, cost function often called path metric is constructed by combining (causal) path metric of the visited nodes and (noncausal) path metric generated from rough estimate of unvisited nodes.
When the generated cost function which corresponds to the sum of two path metrics is greater than the deliberately designed threshold, the search path has little hope to survive in the end and hence is immediately pruned.
This idea can be nicely integrated into the MMP algorithm for bringing further reduction in complexity.
Another way is to trace the single greedy path (which is equivalent to the path of OMP) initially and then initiate the MMP operations afterwards.
In fact, since OMP works pretty well in the early iterations (see Fig. \ref{fig:Candidates}), one can defer the branching operation of MMP.
Noting that the complexity savings by the pruning is pronounced in early layers of the search tree, one can expect that this hybrid strategy will alleviate the computational burden of MMP substantially.
We leave these interesting explorations for our future work.

Finally, we close the paper by quoting the well-known dictum: ``Greed is good" \cite{joeltropp}.
Indeed, we have observed that greedy strategy performs close to optimal one when it is controlled properly.


%

%

\appendices

\section{Proof of Lemma \ref{lem:aL}}
\label{app:aL}

\begin{IEEEproof}
The $\ell_2$-norm of the correlation $\mathbf{\Phi}_{F_L}' \mathbf{r}^{k-1}_i$ is expressed as
\ifCLASSOPTIONonecolumn
\begin{IEEEeqnarray}{rCl}
\left\| \mathbf{\Phi}_{F_L}' {\mathbf{r}^{k-1}_i} \right\|_2	&=& \left\| \mathbf{\Phi}_{F_L}' \mathbf{P}_{s^{k-1}_i}^\bot \mathbf{y} \right\|_2	  \\
	&=& \left\| \mathbf{\Phi}_{F_L}' \mathbf{P}_{s^{k-1}_i}^\bot \mathbf{\Phi}_{T} \mathbf{x}_{T} \right\|_2	  \\
	&=& \left\| \mathbf{\Phi}_{F_L}' \mathbf{P}_{s^{k-1}_i}^\bot \mathbf{\Phi}_{T-s^{k-1}_i} \mathbf{x}_{T-s^{k-1}_i} \right\|_2	  \\
	&=& \left\| \mathbf{\Phi}_{F_L}' \mathbf{\Phi}_{T -s^{k-1}_i} \mathbf{x}_{T - s^{k-1}_i} - \mathbf{\Phi}_{F_L}' \mathbf{P}_{s^{k-1}_i} \mathbf{\Phi}_{T -s^{k-1}_i} \mathbf{x}_{T - s^{k-1}_i} \right\|_2	  \\
	& \leq & \left\| \mathbf{\Phi}_{F_L}' \mathbf{\Phi}_{T -s^{k-1}_i} \mathbf{x}_{T - s^{k-1}_i} \right\|_2 + \left\| \mathbf{\Phi}_{F_L}' \mathbf{P}_{s^{k-1}_i} \mathbf{\Phi}_{T -s^{k-1}_i} \mathbf{x}_{T - s^{k-1}_i} \right\|_2.	 \label{eq:app1_res1}
\end{IEEEeqnarray}
\else
\begin{IEEEeqnarray}{rCl}
\lefteqn{ \left\| \mathbf{\Phi}_{F_L}' {\mathbf{r}^{k-1}_i} \right\|_2	} \nonumber	\\
	&=& \left\| \mathbf{\Phi}_{F_L}' \mathbf{P}_{s^{k-1}_i}^\bot \mathbf{y} \right\|_2	  \nonumber	\\
	&=& \left\| \mathbf{\Phi}_{F_L}' \mathbf{P}_{s^{k-1}_i}^\bot \mathbf{\Phi}_{T} \mathbf{x}_{T} \right\|_2	 \nonumber	 \\
	&=& \left\| \mathbf{\Phi}_{F_L}' \mathbf{P}_{s^{k-1}_i}^\bot \mathbf{\Phi}_{T -s^{k-1}_i} \mathbf{x}_{T - s^{k-1}_i} \right\|_2	  \nonumber	\\
	&=& \left\| \mathbf{\Phi}_{F_L}' \mathbf{\Phi}_{T -s^{k-1}_i} \mathbf{x}_{T - s^{k-1}_i}  \right.	 \nonumber	 \\
	& & \left. - \mathbf{\Phi}_{F_L}' \mathbf{P}_{s^{k-1}_i} \mathbf{\Phi}_{T -s^{k-1}_i} \mathbf{x}_{T - s^{k-1}_i} \right\|_2	  \nonumber	\\
	& \leq & \left\| \mathbf{\Phi}_{F_L}' \mathbf{\Phi}_{T -s^{k-1}_i} \mathbf{x}_{T - s^{k-1}_i} \right\|_2 \nonumber	 \\
	& & + \left\| \mathbf{\Phi}_{F_L}' \mathbf{P}_{s^{k-1}_i} \mathbf{\Phi}_{T -s^{k-1}_i} \mathbf{x}_{T - s^{k-1}_i} \right\|_2.	 \label{eq:app1_res1}
\end{IEEEeqnarray}
\fi

Since $F_L$ and $T - s^{k-1}_i$ are disjoint ($F_L \cap (T -s^{k-1}_i) = \emptyset $) and also noting that the number of correct indices in $s^{k-1}_i$ is $k-1$ by the hypothesis,
\begin{equation}
|F_L| + |T - s^{k-1}_i| = L+K-(k-1).	\label{rev_eq:105}
\end{equation}
Using this together with Lemma \ref{lem:rip_add}, the first term in the right-hand side of \eqref{eq:app1_res1} becomes
\begin{equation}
\left\| \mathbf{\Phi}_{F_L}' \mathbf{\Phi}_{T -s^{k-1}_i} \mathbf{x}_{T - s^{k-1}_i} \right\|_2 \leq  \delta
_{L+K-k+1} \left\| \mathbf{x}_{T - s^{k-1}_i} \right\|_2 .
\label{eq:app1_1}
\end{equation}
Similarly, noting that $F_L \cap s^{k-1}_i = \emptyset$ and  $|F_L| + | s^{k-1}_i |=L+k-1$, the second term in the right-hand side of \eqref{eq:app1_res1} becomes
\begin{IEEEeqnarray}{rCl}
\ifCLASSOPTIONonecolumn
\left \| \mathbf{\Phi}_{F_L}' \mathbf{P}_{s^{k-1}_i} \mathbf{\Phi}_{T -s^{k-1}_i} \mathbf{x}_{T - s^{k-1}_i} \right \|_2
&\leq &
	\delta_{L+k-1} \left \| \mathbf{\Phi}_{s^{k-1}_i}^{\dagger} \mathbf{\Phi}_{T -s^{k-1}_i} \mathbf{x}_{T - s^{k-1}_i} \right \|_2		\\
\else
\lefteqn{ \left \| \mathbf{\Phi}_{F_L}' \mathbf{P}_{s^{k-1}_i} \mathbf{\Phi}_{T -s^{k-1}_i} \mathbf{x}_{T - s^{k-1}_i} \right \|_2 }	\nonumber	\\
&\leq &
\delta_{L+k-1} \left \| \mathbf{\Phi}_{s^{k-1}_i}^{\dagger} \mathbf{\Phi}_{T -s^{k-1}_i} \mathbf{x}_{T - s^{k-1}_i} \right \|_2		 \nonumber	 \\
\fi
	&=& \delta_{L+k-1}  \!\! \left\| \! \left ( \mathbf{\Phi}_{s^{k-1}_i}' \mathbf{\Phi}_{s^{k-1}_i} \right )^{-1}  \!\!\!\! \mathbf{\Phi}_{s^{k-1}_i}' \mathbf{\Phi}_{T -s^{k-1}_i} \mathbf{x}_{T - s^{k-1}_i} \right\|_2		\nonumber	\\
	& \overset{ (a) }{\leq} & \frac{\delta_{L+k-1}}{1-\delta_{k-1}} \left\| \mathbf{\Phi}_{s^{k-1}_i}' \mathbf{\Phi}_{T -s^{k-1}_i} \mathbf{x}_{T - s^{k-1}_i} \right\|_2		\nonumber	\\
	& \overset{ (b) }{\leq} & \frac{\delta_{L+k-1} \delta_{(k-1)+K-(k-1)}}{1-\delta_{k-1}} \left\| \mathbf{x}_{T - s^{k-1}_i} \right\|_2		 \nonumber	\\
	&=& \frac{\delta_{L+k-1} \delta_{K}}{1-\delta_{k-1}} \left\| \mathbf{x}_{T - s^{k-1}_i} \right\|_2	 \label{eq:app1_5}
\end{IEEEeqnarray}
where (a) and (b) follow from Lemma \ref{lem:rip_cons} and \ref{lem:rip_add}, respectively.
%
%

Using \eqref{eq:app1_res1}, \eqref{eq:app1_1}, and \eqref{eq:app1_5}, we have
\begin{equation}
\left\| \mathbf{\Phi }_{F_L}' \mathbf{r}^{k-1}_i \right\|_2
\leq
\left( \delta _{L+K-k+1} + \frac{\delta _{L+k-1} \delta_{K}} {1 - \delta_{k-1}} \right ) \left \|
\mathbf{x}_{T - s^{k-1}_i} \right \|_2.	\label{eq:app1_6}
\end{equation}
%
Further, recalling that
\begin{IEEEeqnarray}{rCl}
\left\| \mathbf{\Phi }_{F_L}' \mathbf{r}^{k-1}_i \right\|_2^2 = \sum\limits_{j = 1}^L { \left | \left< \mathbf{\phi}_{f_j}, \mathbf{r}^{k-1}_i \right> \right |^2} = \sum\limits_{j = 1}^L { \left ( \alpha_j^k \right )^2 } \nonumber
\end{IEEEeqnarray}
we have,
\begin{IEEEeqnarray}{rCl}
\left\| \mathbf{\Phi }_{F_L}' \mathbf{r}^{k-1}_i \right\|_2 &\geq& \frac{1}{\sqrt L }\sum\limits_{i = 1}^L { \alpha_i^k } 	 \label{eq:app1_norm_ineq1} \\
	&\geq& \frac{1} {\sqrt L } L \alpha_L^k   = \sqrt L  \alpha_L^k		\label{eq:app1_norm_ineq}
\end{IEEEeqnarray}
where \eqref{eq:app1_norm_ineq1} follows from the norm inequality $\left ( \left\| \mathbf{z}\right\|_1 \leq \sqrt{\left\| \mathbf{z} \right\|_0} \left\|\mathbf{z} \right\|_2 \right )$ and \eqref{eq:app1_norm_ineq} is because $\alpha_1^k \geq  \alpha_2^k \geq \cdots \geq \alpha_L^k $.
%
Combining \eqref{eq:app1_6} and \eqref{eq:app1_norm_ineq}, we have
\begin{equation}
\left( \delta _{L + K - k+1} + \frac{\delta _{L+k-1} \delta_{K}} {1 - \delta_{k-1}} \right ) \left \|
\mathbf{x}_{T - s^{k-1}_i} \right \|_2 \geq \sqrt L \alpha_L^k,
\end{equation}
and hence
\begin{equation}
\alpha_L^k \leq \left( \delta _{L+K-k+1} + \frac{\delta _{L+k-1} \delta_{K}} {1 - \delta_{k-1}} \right ) \frac{\left \| \mathbf{x}_{T - s^{k-1}_i} \right \|_2 }{\sqrt{L}}.
\end{equation}
\end{IEEEproof}

\section{Proof of Lemma \ref{lem:b1}}
\label{app:b1}
\begin{IEEEproof}
Since $\beta_1^k$ is the largest correlation in magnitude between $\mathbf{r}^{k-1}_i$ and $\left \{ \phi_j \right \}_{j \in T - s^{k-1}_i}$ $\left ( \left|\left<\mathbf{\phi}_{\varphi_j}, \mathbf{r}^{k-1}_i\right>\right| \right )$\footnote{$\varphi_j = \arg \mathop {\max} \limits_{ u \in \left( T-s^{k-1}_i \right)\setminus \{\varphi_1, \ldots, \varphi_{j-1}\}} \left|\left <\mathbf{\phi}_u, \mathbf{r}^{k-1}_i \right> \right|$}, it is clear that
\begin{equation}
	\beta_1^k \geq \left | \left < \phi_j, \mathbf{r}^{k-1}_i \right >\right |
\end{equation}
for all $j \in T- s^{k-1}_i$. Noting that $|T - s_i^{k-1}| = K - (k -1)$, we have
%
%
\begin{IEEEeqnarray}{rCl}
	\beta_1^k & > & \frac{1}{\sqrt{K-(k-1)}}\left \| \mathbf{\Phi}'_{T-s^{k-1}_i} \mathbf{r}^{k-1}_i \right \|_2		 \label{rev_eq:120}	\\
			&= & \frac{1}{\sqrt{K-k+1}}\left \| \mathbf{\Phi}'_{T-s^{k-1}_i} \mathbf{P}_{s^{k-1}_i}^ \bot \mathbf{\Phi} \mathbf{x} \right \|_2		\label{eq:app2_1}
\end{IEEEeqnarray}
where \eqref{eq:app2_1} follows from $\mathbf{r}^{k-1}_i = \mathbf{y} - \mathbf{\Phi}_{s^{k-1}_i} \mathbf{\Phi}_{s^{k-1}_i}^\dag \mathbf{y} = \mathbf{P}_{s^{k-1}_i}^ \bot {\mathbf{y}}$.
Using the triangle inequality,
\ifCLASSOPTIONonecolumn
\begin{IEEEeqnarray}{rCl}
\beta_1^k
&\geq & \frac{1}{\sqrt{K-k+1}}\left\| \mathbf{\Phi }_{T - s^{k-1}_i}' \mathbf{P}_{s^{k-1}_i}^ \bot \mathbf{\Phi }_{T - s^{k-1}_i} \mathbf{x}_{T - s^{k-1}_i} \right\|_2  \\
&\geq & \frac{ \left\| \mathbf{\Phi}_{T - s^{k-1}_i}' \mathbf{\Phi }_{T - s^{k-1}_i} \mathbf{x}_{T - s^{k-1}_i} \right\|_2 - \left\| \mathbf{\Phi}_{T - s^{k-1}_i}' \mathbf{P}_{s^{k-1}_i} \mathbf{\Phi }_{T - s^{k-1}_i} \mathbf{x}_{T - s^{k-1}_i} \right\|_2 }{\sqrt{K-k+1}}.	\label{eq:app2_7}
\end{IEEEeqnarray}
\else
\begin{IEEEeqnarray}{rCl}
\beta_1^k
&\geq& \frac{\left\| \mathbf{\Phi }_{T - s^{k-1}_i}' \mathbf{P}_{s^{k-1}_i}^ \bot \mathbf{\Phi }_{T - s^{k-1}_i} \mathbf{x}_{T - s^{k-1}_i} \right\|_2}{\sqrt{K-k+1}}  \\
&\geq & \frac{ \left\| \mathbf{\Phi}_{T - s^{k-1}_i}' \mathbf{\Phi }_{T - s^{k-1}_i} \mathbf{x}_{T - s^{k-1}_i} \right\|_2}{\sqrt{K-k+1}}	\nonumber	\\
& & -\frac{ \left\| \mathbf{\Phi}_{T - s^{k-1}_i}' \mathbf{P}_{s^{k-1}_i} \mathbf{\Phi }_{T - s^{k-1}_i} \mathbf{x}_{T - s^{k-1}_i} \right\|_2 }{\sqrt{K-k+1}}.	\label{eq:app2_7}
\end{IEEEeqnarray}
\fi
%
Since $\left | T-s^{k-1}_i \right | = K-(k-1)$, using Lemma \ref{lem:rip_cons}, we have
\begin{equation}
	\left\| \mathbf{\Phi}_{T - s^{k-1}_i}' \mathbf{\Phi }_{T - s^{k-1}_i} \mathbf{x}_{T - s^{k-1}_i} \right\|_2 \geq \left( 1-\delta_{K-k+1}\right) \left \| \mathbf{x}_{T-s^{k-1}_i} \right \|_2	 \label{eq:app2_8}
\end{equation}
and also
\begin{IEEEeqnarray}{rCl}
\lefteqn{ \left\| \mathbf{\Phi}_{T - s^{k-1}_i}' \mathbf{P}_{s^{k-1}_i} \mathbf{\Phi }_{T - s^{k-1}_i} \mathbf{x}_{T - s^{k-1}_i} \right\|_2 }	\nonumber	\\
&=& \!\!\!
\left\| \mathbf{\Phi}_{T - s^{k-1}_i}' \mathbf{\Phi}_{s^{k-1}_i} \left ( \!\! \mathbf{\Phi}_{s^{k-1}_i}' \mathbf{\Phi}_{s^{k-1}_i} \!\! \right )^{-1} \!\!\!\! \mathbf{\Phi}_{s^{k-1}_i}' \mathbf{\Phi }_{T - s^{k-1}_i} \mathbf{x}_{T - s^{k-1}_i} \right\|_2	\nonumber	\\
&\overset{ (a) }{\leq} & \!
\delta_{k-1+K-(k-1)} \left\| \left ( \!\! \mathbf{\Phi}_{s^{k-1}_i}' \mathbf{\Phi}_{s^{k-1}_i} \!\! \right )^{-1} \!\!\!\! \mathbf{\Phi}_{s^{k-1}_i}' \mathbf{\Phi }_{T - s^{k-1}_i} \mathbf{x}_{T - s^{k-1}_i} \right\|_2	 \nonumber \\
&\overset{ (b) }{\leq} &
\frac{\delta_{K}}{1-\delta_{k-1}} \left\| \mathbf{\Phi}_{s^{k-1}_i}' \mathbf{\Phi }_{T - s^{k-1}_i} \mathbf{x}_{T - s^{k-1}_i} \right\|_2	\nonumber	\\
&\overset{ (c) }{\leq} &
\frac{\delta_{K}\delta_{(k-1)+K-(k-1)}}{1-\delta_{k-1}} \left\| \mathbf{x}_{T - s^{k-1}_i} \right\|_2	 \label{eq:app2_6}.
\end{IEEEeqnarray}
where (a) and (c) follow from Lemma \ref{lem:rip_add}, and (b) follows from Lemma \ref{lem:rip_cons}.
Finally, by combining \eqref{eq:app2_7}, \eqref{eq:app2_8} and \eqref{eq:app2_6}, we have
\begin{equation}
\beta_1^k
\geq
\left( 1- \delta_{K-k+1}-\frac{\delta_{K}^2}{1-\delta_{k-1}} \right) \frac{\left\| \mathbf{x}_{T - s^{k-1}_i} \right\|_2}{\sqrt{K-k+1}}.
\end{equation}
\end{IEEEproof}

\section{Proof of Lemma \ref{lem:aL_noisy}}
\label{app:aL_n}

\begin{IEEEproof}
Using the triangle inequality, we have
\ifCLASSOPTIONonecolumn
\begin{IEEEeqnarray}{rCl}
\left\| \mathbf{\Phi}_{F_L}' {\mathbf{r}^{k-1}_i} \right\|_2	&=& \left\| \mathbf{\Phi}_{F_L}' \mathbf{P}_{s^{k-1}_i}^\bot \left (\mathbf{\Phi}_T \mathbf{x}_T + \mathbf{v} \right ) \right\|_2	  \\
	&\leq& \left\| \mathbf{\Phi}_{F_L}' \mathbf{P}_{s^{k-1}_i}^\bot \mathbf{\Phi}_{T -s^{k-1}_i} \mathbf{x}_{T - s^{k-1}_i} \right\|_2	+ \left\| \mathbf{\Phi}_{F_L}' \mathbf{P}_{s^{k-1}_i}^\bot \mathbf{v} \right\|_2	 \label{eq:n_gen_init}.
\end{IEEEeqnarray}
\else
\begin{IEEEeqnarray}{rCl}
\left\| \mathbf{\Phi}_{F_L}' {\mathbf{r}^{k-1}_i} \right\|_2	&=& \left\| \mathbf{\Phi}_{F_L}' \mathbf{P}_{s^{k-1}_i}^\bot \left (\mathbf{\Phi}_T \mathbf{x}_T + \mathbf{v} \right ) \right\|_2	\nonumber  \\
	&\leq& \left\| \mathbf{\Phi}_{F_L}' \mathbf{P}_{s^{k-1}_i}^\bot \mathbf{\Phi}_{T -s^{k-1}_i} \mathbf{x}_{T - s^{k-1}_i} \right\|_2	\nonumber	\\
	& & + \left\| \mathbf{\Phi}_{F_L}' \mathbf{P}_{s^{k-1}_i}^\bot \mathbf{v} \right\|_2	 \label{eq:n_gen_init}.
\end{IEEEeqnarray}
\fi
%
Using \eqref{eq:app1_6} in Appendix \ref{app:aL}, we have
%
\ifCLASSOPTIONonecolumn
\begin{equation}
\left\| \mathbf{\Phi}_{F_L}' \mathbf{P}_{s^{k-1}_i}^\bot \mathbf{\Phi}_{T -s^{k-1}_i} \mathbf{x}_{T - s^{k-1}_i} \right\|_2
\leq
\left( \delta _{L+K-k+1} + \frac{\delta _{L+k-1} \delta_{K}} {1 - \delta_{k-1}} \right ) \left \|
\mathbf{x}_{T - s^{k-1}_i} \right \|_2. \label{eq:n_gen_an1}
\end{equation}
\else
\begin{IEEEeqnarray}{rCl}
\lefteqn{ \left\| \mathbf{\Phi}_{F_L}' \mathbf{P}_{s^{k-1}_i}^\bot \mathbf{\Phi}_{T -s^{k-1}_i} \mathbf{x}_{T - s^{k-1}_i} \right\|_2	}	\nonumber	\\
& \leq & \left( \delta _{L+K-k+1} + \frac{\delta _{L+k-1} \delta_{K}} {1 - \delta_{k-1}} \right ) \left \|
\mathbf{x}_{T - s^{k-1}_i} \right \|_2. \label{eq:n_gen_an1}
\end{IEEEeqnarray}
\fi
%
In addition, we have
%
\begin{IEEEeqnarray}{rCl}
\left\| \mathbf{\Phi}_{F_L}' \mathbf{P}_{s^{k-1}_i}^\bot \mathbf{v} \right\|_2
	& \leq & \left\| \mathbf{\Phi}_{F_L}' \right\|_2 \left\| \mathbf{P}_{s^{k-1}_i}^\bot \mathbf{v} \right\|_2	\nonumber	 \\
	& \overset{ (a) }{\leq} & \sqrt{ 1+\delta_L } \left\| \mathbf{P}_{s^{k-1}_i}^\bot \mathbf{v} \right\|_2		 \nonumber	 \\
	& \leq & \sqrt{ 1+\delta_L } \left\| \mathbf{v} \right\|_2	\label{eq:n_gen_an2}
\end{IEEEeqnarray}
where (a) is from Lemma \ref{lem:matrix_up}.

Plugging \eqref{eq:n_gen_an1} and \eqref{eq:n_gen_an2} into \eqref{eq:n_gen_init}, we have
\ifCLASSOPTIONonecolumn
\begin{equation}
\left\| \mathbf{\Phi}_{F_L}' {\mathbf{r}^{k-1}_i} \right\|_2	\leq
\left( \delta _{L+K-k+1} + \frac{\delta _{L+k-1} \delta_{K}} {1 - \delta_{k-1}} \right ) \left \|
\mathbf{x}_{T - s^{k-1}_i} \right \|_2
+\sqrt{ 1+\delta_L } \left\| \mathbf{v} \right\|_2.	\label{eq:n_gen_an3}
\end{equation}
\else
\begin{IEEEeqnarray}{rCl}
\lefteqn{ \left\| \mathbf{\Phi}_{F_L}' {\mathbf{r}^{k-1}_i} \right\|_2	\leq }	\nonumber	\\
& & \left( \! \delta _{L+K-k+1} + \frac{\delta _{L+k-1} \delta_{K}} {1 - \delta_{k-1}} \!  \right ) \!\! \left \|\mathbf{x}_{T - s^{k-1}_i} \!  \right \|_2	\!\!  + \sqrt{ 1+\delta_L } \left\| \mathbf{v} \right\|_2.	\IEEEeqnarraynumspace \label{eq:n_gen_an3}
\end{IEEEeqnarray}
\fi
%
Further, using the norm inequality (see \eqref{eq:app1_norm_ineq}), we have
\begin{equation}
\left\| \mathbf{\Phi }_{F_L}' \mathbf{r}^{k-1}_i \right\|_2
	\geq \frac{1}{\sqrt L }\sum\limits_{i = 1}^L { \alpha_i^k }
	\geq \frac{1} {\sqrt L } L \alpha_L^k   = \sqrt L  \alpha_L^k.
\label{eq:app1_n_norm_ineq}
\end{equation}
%
Combining \eqref{eq:n_gen_an3} and \eqref{eq:app1_n_norm_ineq}, we have
\ifCLASSOPTIONonecolumn
\begin{equation}
\left( \delta _{L + K - k+1} + \frac{\delta _{L+k-1} \delta_{K}} {1 - \delta_{k-1}} \right ) \left \| \mathbf{x}_{T - s^{k-1}_i} \right \|_2
+\sqrt{ 1+\delta_L } \left\| \mathbf{v} \right\|_2
\geq \sqrt L \alpha_L^k,
\end{equation}
\else
\begin{IEEEeqnarray}{rCl}
\left( \delta _{L + K - k+1} + \frac{\delta _{L+k-1} \delta_{K}} {1 - \delta_{k-1}} \right ) \left \| \mathbf{x}_{T - s^{k-1}_i} \right \|_2	+\sqrt{ 1+\delta_L } \left\| \mathbf{v} \right\|_2 	\nonumber	\\
\geq  \sqrt L \alpha_L^k,	\IEEEeqnarraynumspace
\end{IEEEeqnarray}
\fi
and hence
\ifCLASSOPTIONonecolumn
\begin{equation}
\alpha_L^k \leq \left( \delta _{L + K - k+1} + \frac{\delta _{L+k-1} \delta_{K}} {1 - \delta_{k-1}} \right ) \frac{\left \| \mathbf{x}_{T - s^{k-1}_i} \right \|_2}{\sqrt{L}}
+ \frac{\sqrt{ 1+\delta_L } \left\| \mathbf{v} \right\|_2}{\sqrt{L}}.
\end{equation}
\else
\begin{IEEEeqnarray}{rCl}
\alpha_L^k & \leq & \left( \delta _{L + K - k+1} + \frac{\delta _{L+k-1} \delta_{K}} {1 - \delta_{k-1}} \right )	\frac{\left \| \mathbf{x}_{T - s^{k-1}_i} \right \|_2}{\sqrt{L}}  \nonumber	\\
& & + \frac{\sqrt{ 1+\delta_L } \left\| \mathbf{v} \right\|_2}{\sqrt{L}}.
\end{IEEEeqnarray}
\fi

\end{IEEEproof}

\section{Proof of Lemma \ref{lem:b1_noisy}}
\label{app:b1_n}

\begin{IEEEproof}
Using the definition of $\beta_1^k$ (see \eqref{eq:def_b1} and \eqref{eq:app2_7}), we have
\begin{equation}
	\beta_1^k
		 \geq \frac{1}{\sqrt{K-(k-1)}}\left \| \mathbf{\Phi}'_{T-s^{k-1}_i} \mathbf{r}^{k-1}_i \right \|_2
\end{equation}
and
\ifCLASSOPTIONonecolumn
\begin{IEEEeqnarray}{rCl}
	\lefteqn{\frac{1}{\sqrt{K-(k-1)}}\left \| \mathbf{\Phi}'_{T-s^{k-1}_i} \mathbf{r}^{k-1}_i \right \|_2} \\
			& = & \frac{1}{\sqrt{K-k+1}}\left \| \mathbf{\Phi}'_{T-s^{k-1}_i} \mathbf{P}_{s^{k-1}_i}^ \bot \mathbf{y} \right \|_2		 \\
			& = & \frac{1}{\sqrt{K-k+1}}\left \| \mathbf{\Phi}'_{T-s^{k-1}_i} \mathbf{P}_{s^{k-1}_i}^ \bot \left (\mathbf{\Phi}_T \mathbf{x}_T + \mathbf{v} \right ) \right \|_2		\\
			& \geq & \frac{1}{\sqrt{K-k+1}}\left \| \mathbf{\Phi}'_{T-s^{k-1}_i} \mathbf{P}_{s^{k-1}_i}^ \bot \mathbf{\Phi}_T \mathbf{x}_T \right \|_2
			-
			\frac{1}{\sqrt{K-k+1}}\left \| \mathbf{\Phi}'_{T-s^{k-1}_i} \mathbf{P}_{s^{k-1}_i}^ \bot \mathbf{v} \right \|_2.
\end{IEEEeqnarray}
\else
\begin{IEEEeqnarray}{rCl}
	\lefteqn{\frac{1}{\sqrt{K-(k-1)}}\left \| \mathbf{\Phi}'_{T-s^{k-1}_i} \mathbf{r}^{k-1}_i \right \|_2} \nonumber	\\
			& = & \frac{1}{\sqrt{K-k+1}}\left \| \mathbf{\Phi}'_{T-s^{k-1}_i} \mathbf{P}_{s^{k-1}_i}^ \bot \mathbf{y} \right \|_2	 \nonumber \\
			& = & \frac{1}{\sqrt{K-k+1}}\left \| \mathbf{\Phi}'_{T-s^{k-1}_i} \mathbf{P}_{s^{k-1}_i}^ \bot \left (\mathbf{\Phi}_T \mathbf{x}_T + \mathbf{v} \right ) \right \|_2	\nonumber	\\
			& \geq & \frac{1}{\sqrt{K-k+1}}\left \| \mathbf{\Phi}'_{T-s^{k-1}_i} \mathbf{P}_{s^{k-1}_i}^ \bot \mathbf{\Phi}_T \mathbf{x}_T \right \|_2	\nonumber	\\
			& & -
			\frac{1}{\sqrt{K-k+1}}\left \| \mathbf{\Phi}'_{T-s^{k-1}_i} \mathbf{P}_{s^{k-1}_i}^ \bot \mathbf{v} \right \|_2.
\end{IEEEeqnarray}
\fi
%
Recalling from \eqref{eq:app2_8} and \eqref{eq:app2_6} in Appendix \ref{app:b1}, we further have
\ifCLASSOPTIONonecolumn
\begin{equation}
\left \| \mathbf{\Phi}'_{T-s^{k-1}_i} \mathbf{P}_{s^{k-1}_i}^ \bot \mathbf{\Phi}_T \mathbf{x}_T \right \|_2
\geq
\left( 1- \delta_{K-k+1}-\frac{\delta_{K}^2}{1-\delta_{k-1}} \right) \left\| \mathbf{x}_{T - s^{k-1}_i} \right\|_2 . \label{eq:n_gen_b11}
\end{equation}
\else
\begin{IEEEeqnarray}{rCl}
\lefteqn{ \left \| \mathbf{\Phi}'_{T-s^{k-1}_i} \mathbf{P}_{s^{k-1}_i}^ \bot \mathbf{\Phi}_T \mathbf{x}_T \right \|_2 } 	 \nonumber	\\
&\geq &  \left( 1- \delta_{K-k+1}- \frac{\delta_{K}^2}{1-\delta_{k-1}} \right) \left\| \mathbf{x}_{T - s^{k-1}_i} \right\|_2 . \label{eq:n_gen_b11}
\end{IEEEeqnarray}
\fi
Next, the upper bound of $\left \| \mathbf{\Phi}'_{T-s^{k-1}_i} \mathbf{P}_{s^{k-1}_i}^ \bot \mathbf{v} \right \|_2$ becomes
\begin{IEEEeqnarray}{rCl}
\left \| \mathbf{\Phi}'_{T-s^{k-1}_i} \mathbf{P}_{s^{k-1}_i}^ \bot \mathbf{v} \right \|_2
	& \leq & \left \| \mathbf{\Phi}'_{T-s^{k-1}_i} \right \|_2 \left \| \mathbf{P}_{s^{k-1}_i}^ \bot \mathbf{v} \right \|_2		 \nonumber \\
	& \overset{ (a) }{\leq} & \sqrt{1+\delta_{K-k+1}} \left \| \mathbf{P}_{s^{k-1}_i}^ \bot \mathbf{v} \right \|_2	\nonumber	 \\
	& \leq & \sqrt{1+\delta_{K-k+1}} \left \| \mathbf{v} \right \|_2 \label{eq:n_gen_b12}
\end{IEEEeqnarray}
where (a) follows from Lemma \ref{lem:matrix_up}.
Combining \eqref{eq:n_gen_b11} and \eqref{eq:n_gen_b12}, we get the desired result.
\end{IEEEproof}

\section{The lower bound of residual}
\label{app:lb_resi}
\begin{IEEEproof}
Let $\Gamma$ be the set of $K$ indices, then we have
\begin{IEEEeqnarray}{rCl}
\| \mathbf{r}_{\Gamma} \|_2^2 &=& \|\mathbf{P}_{\Gamma}^{\bot}\mathbf{y}\|_2^2		\nonumber \\
	&=& \|\mathbf{P}_{\Gamma}^{\bot}(\mathbf{\Phi}_T\mathbf{x}_T + \mathbf{v})\|_2^2		\nonumber \\
	&\geq& \|\mathbf{P}_{\Gamma}^{\bot}\mathbf{\Phi}_T\mathbf{x}_T\|_2^2 - \|\mathbf{P}_{\Gamma}^{\bot} \mathbf{v}\|_2^2		 \nonumber \\
	&\geq& \|\mathbf{P}_{\Gamma}^{\bot}\mathbf{\Phi}_T\mathbf{x}_T\|_2^2 - \|\mathbf{v}\|_2^2		 \label{app:lwr_rl_m}.
\end{IEEEeqnarray}
Furthermore,
\begin{IEEEeqnarray}{rCl}
\ifCLASSOPTIONonecolumn
\|\mathbf{P}_{\Gamma}^{\bot}\mathbf{\Phi}_T\mathbf{x}_T\|_2^2 &=& \|\mathbf{P}_{\Gamma}^{\bot}\mathbf{\Phi}_{T-\Gamma}\mathbf{x}_{T-\Gamma}\|_2^2	\nonumber \\
\else
\lefteqn{ \|\mathbf{P}_{\Gamma}^{\bot}\mathbf{\Phi}_T\mathbf{x}_T\|_2^2 } \nonumber	\\
	&=& \|\mathbf{P}_{\Gamma}^{\bot}\mathbf{\Phi}_{T-\Gamma}\mathbf{x}_{T-\Gamma}\|_2^2	\nonumber \\
\fi
	&=& \|\mathbf{\Phi}_{T-\Gamma}\mathbf{x}_{T-\Gamma}-\mathbf{P}_{\Gamma}\mathbf{\Phi}_{T-\Gamma}\mathbf{x}_{T-\Gamma}\|_2^2		 \nonumber \\
	&\geq& \|\mathbf{\Phi}_{T-\Gamma}\mathbf{x}_{T-\Gamma}\|_2^2-\|\mathbf{P}_{\Gamma}\mathbf{\Phi}_{T-\Gamma}\mathbf{x}_{T-\Gamma}\|_2^2	 \nonumber \\
	&\geq& (1-\delta_{|T-\Gamma|})\|\mathbf{x}_{T-\Gamma}\|_2^2-\|\mathbf{P}_{\Gamma}\mathbf{\Phi}_{T-\Gamma}\mathbf{x}_{T-\Gamma}\|_2^2	 \label{app:lwr_pi_m},
\end{IEEEeqnarray}
and
\begin{IEEEeqnarray}{rCl}
\ifCLASSOPTIONonecolumn
	\|\mathbf{P}_{\Gamma}\mathbf{\Phi}_{T-\Gamma}\mathbf{x}_{T-\Gamma}\|_2^2 &=& \|\mathbf{\Phi}_{\Gamma} \left( \mathbf{\Phi}_{\Gamma}'\mathbf{\Phi}_{\Gamma} \right)^{-1}\mathbf{\Phi}_{\Gamma}'\mathbf{\Phi}_{T-\Gamma}\mathbf{x}_{T-\Gamma}\|_2^2	\nonumber \\
\else
\lefteqn{ \|\mathbf{P}_{\Gamma}\mathbf{\Phi}_{T-\Gamma}\mathbf{x}_{T-\Gamma}\|_2^2 } \nonumber	\\
	&=& \|\mathbf{\Phi}_{\Gamma} \left( \mathbf{\Phi}_{\Gamma}'\mathbf{\Phi}_{\Gamma} \right)^{-1}\mathbf{\Phi}_{\Gamma}'\mathbf{\Phi}_{T-\Gamma}\mathbf{x}_{T-\Gamma}\|_2^2	\nonumber \\
\fi
	&\overset{ (a) }{\leq}& (1+\delta_{|\Gamma|})\| \left( \mathbf{\Phi}_{\Gamma}'\mathbf{\Phi}_{\Gamma} \right)^{-1}\mathbf{\Phi}_{\Gamma}'\mathbf{\Phi}_{T-\Gamma}\mathbf{x}_{T-\Gamma}\|_2^2		 \nonumber	\\
	&\overset{ (b) }{\leq}& \frac{1+\delta_{|\Gamma|}}{(1-\delta_{|\Gamma|})^2}\| \mathbf{\Phi}_{\Gamma}'\mathbf{\Phi}_{T-\Gamma}\mathbf{x}_{T-\Gamma}\|_2^2		\nonumber	\\
	&\overset{ (c) }{\leq}& \frac{(1+\delta_{|\Gamma|})\delta_{|\Gamma|+|T-\Gamma|}^2}{(1-\delta_{|\Gamma|})^2}\|\mathbf{x}_{T-\Gamma}\|_2^2	 \label{app:lwr_p}
\end{IEEEeqnarray}
where (a) is from the definition of RIP, (b) and (c) are from Lemma \ref{lem:rip_cons} and \ref{lem:rip_add}, respectively.
Combining \eqref{app:lwr_pi_m} and \eqref{app:lwr_p}, 
we have
%
\ifCLASSOPTIONonecolumn
\begin{IEEEeqnarray}{rCl}
	\|\mathbf{P}_{\Gamma}^{\bot}\mathbf{\Phi}_T\mathbf{x}_T\|_2^2 &\geq& (1-\delta_{|T-\Gamma|})\|\mathbf{x}_{T-\Gamma}\|_2^2 -
			 \frac{(1+\delta_{|\Gamma|})\delta_{|\Gamma|+|T-\Gamma|}^2}{(1-\delta_{|\Gamma|})^2}\|\mathbf{x}_{T-\Gamma}\|_2^2		 \\
	&=& \left( (1-\delta_{|T-\Gamma|})-
			\frac{(1+\delta_{|\Gamma|})\delta_{|\Gamma|+|T-\Gamma|}^2}{(1-\delta_{|\Gamma|})^2} \right) \|\mathbf{x}_{T-\Gamma}\|_2^2	\label{app:lwr_pi}.
\end{IEEEeqnarray}
\else
\begin{IEEEeqnarray}{rCl}
	\lefteqn{ \|\mathbf{P}_{\Gamma}^{\bot}\mathbf{\Phi}_T\mathbf{x}_T\|_2^2 }	\nonumber	\\
	&\geq & (1-\delta_{|T-\Gamma|})\|\mathbf{x}_{T-\Gamma}\|_2^2 - \frac{(1+\delta_{|\Gamma|})\delta_{|\Gamma|+|T-\Gamma|}^2}{(1-\delta_{|\Gamma|})^2}\|\mathbf{x}_{T-\Gamma}\|_2^2 \nonumber	\\
	&=& \left( (1-\delta_{|T-\Gamma|})-
			\frac{(1+\delta_{|\Gamma|})\delta_{|\Gamma|+|T-\Gamma|}^2}{(1-\delta_{|\Gamma|})^2} \right) \|\mathbf{x}_{T-\Gamma}\|_2^2	\label{app:lwr_pi}.
\end{IEEEeqnarray}
\fi
%
Finally, using \eqref{app:lwr_rl_m} and \eqref{app:lwr_pi}, 
we have
\ifCLASSOPTIONonecolumn
\begin{equation}
\| \mathbf{r}_{\Gamma} \|_2^2 \geq \left( (1-\delta_{|T-\Gamma|})-
			\frac{(1+\delta_{|\Gamma|})\delta_{|\Gamma|+|T-\Gamma|}^2}{(1-\delta_{|\Gamma|})^2} \right) \|\mathbf{x}_{T-\Gamma}\|_2^2 - \|\mathbf{v}\|_2^2.
\end{equation}
\else
\begin{IEEEeqnarray}{rCl}
\| \mathbf{r}_{\Gamma} \|_2^2
&\geq& \left( (1-\delta_{|T-\Gamma|})-
			\frac{(1+\delta_{|\Gamma|})\delta_{|\Gamma|+|T-\Gamma|}^2}{(1-\delta_{|\Gamma|})^2} \right) \|\mathbf{x}_{T-\Gamma}\|_2^2 \nonumber	\\
& &- \|\mathbf{v}\|_2^2.
\end{IEEEeqnarray}
\fi
\end{IEEEproof}



%



\section*{Acknowledgment}
The authors would like to thank the anonymous reviewers and an associate editor Prof. O. Milenkovic for their valuable comments and suggestions that improved the quality of the paper.
In particular, reviewers pointed out the mistakes of proofs in Section~\ref{sec:noisy_anal}.

\ifCLASSOPTIONcaptionsoff
  \newpage
\fi

\begin{IEEEbiographynophoto}{Suhyuk Kwon (S' 11) }
received the B.S. and M.S. degrees in the School of Information and Communication from Korea University, Seoul, Korea, in 2008 and 2010, where he is currently working toward the Ph.D. degree. His research interests include compressive sensing, signal processing, and information theory.
\end{IEEEbiographynophoto}

\begin{IEEEbiographynophoto}{Jian Wang (S' 11) }
received the B.S. degree in material engineering and the M.S. degree in information and communication engineering from Harbin Institute of Technology, China, in 2006 and 2009, respectively, and the Ph.D. degree in electrical and computer engineering from Korea University, Korea in 2013.
Currently he holds a postdoctoral research associate position in Dept. of Statistics at Rutgers University, NJ. His research interests include compressed sensing, signal processing in wireless communications, and statistical learning.
\end{IEEEbiographynophoto}

\begin{IEEEbiographynophoto}{Byonghyo Shim (SM' 09)}
received the B.S. and M.S. degrees in control and instrumentation engineering (currently electrical engineering) from Seoul National University, Korea, in 1995 and 1997, respectively and the M.S. degree in Mathematics and the Ph.D. degree in Electrical and Computer Engineering from the University of Illinois at Urbana-Champaign, in 2004 and 2005, respectively.

From 1997 and 2000, he was with the Department of Electronics Engineering at the Korean Air Force Academy as an Officer (First Lieutenant) and an Academic Full-time Instructor. From 2005 to 2007, he was with Qualcomm Inc., San Diego, CA., working on CDMA systems. Since September 2007, he has been with the School of Information and Communication, Korea University, where he is presently an Associate Professor. His research interests include statistical signal processing, compressive sensing, wireless communications, and information theory.

Dr. Shim was the recipient of the 2005 M. E. Van Valkenburg Research Award from the Electrical and Computer Engineering Department of the University of Illinois and 2010 Hadong Young Engineer Award from IEEK.
He is currently an associate editor of IEEE Wireless Communications Letters and a guest editor of IEEE Journal on Selected Areas in Communications (JSAC).
\end{IEEEbiographynophoto}

\end{document}